# Label-free virtual Hematoxylin and Eosin (H&E) staining using second generation Photoacoustic Remote Sensing (PARS)

Benjamin R. Ecclestone,[1,2] Kevan Bell,[1,2] Sarah Sparkes,[1] Deepak Dinakaran,[3] John R. Mackey,[3] Parsin Haji Reza[1,*]

[1] PhotoMedicine Labs, Department of System Design Engineering, University of Waterloo, 200 University Ave W, Waterloo, ON, N2L 3G1, Canada
[2] IllumiSonics Inc, Department of System Design Engineering, University of Waterloo, 200 University Ave W, Waterloo, ON N2L 3G1, Canada
[3] Cross Cancer Institute, Department of Oncology, University of Alberta, 116 St & 85 Ave, Edmonton, Alberta, T6G 2V1, Canada

*Corresponding Author:* phajireza@uwaterloo.ca

## Abstract

In the past decades, absorption modalities have emerged as powerful tools for label-free functional and structural imaging of cells and tissues. Many biomolecules present unique absorption spectra providing chromophore-specific information on properties such as chemical bonding, and sample composition. As chromophores absorb photons the absorbed energy is emitted as photons (radiative relaxation) or converted to heat and under specific conditions pressure (non-radiative relaxation). Modalities like fluorescence microscopy may capture radiative relaxation to provide contrast, while modalities like photoacoustic microscopy may leverage non-radiative heat and pressures. Here we show an all-optical non-contact total-absorption photoacoustic remote sensing (TA-PARS) microscope, which can capture both radiative and non-radiative absorption effects in a single acquisition. The TA-PARS yields an absorption metric proposed as the quantum efficiency ratio (QER), which visualizes a biomolecules proportional radiative and non-radiative absorption response. The TA-PARS provides label-free visualization of a range of biomolecules enabling convincing analogues to traditional histochemical staining of tissues, effectively providing label-free Hematoxylin and Eosin (H&E)-like visualizations. These findings represent the establishment of an effective all-optical non-contact total-absorption microscope for label-free inspection of biological media.

## Introduction

Microscopic optical inspection techniques have enabled a multitude of breakthroughs in biological research and clinical diagnostics over the past few centuries. From cytological assessment of mitotic cells in oncology,[1] to structural imaging of neurons in biological research,[2] optical microscopy has provided valuable insights into the composition, structure, and function of tissues and cells.[3] Modern optical techniques broadly leverage scattering and absorption events to deliver visualizations in biological media, where each mechanism imparts different characteristics to the respective modalities.

Scattering-based modalities such as OCT,[4] and darkfield microscopy,[5] leverage scattering interactions to provide visualization of structural composition.[6] Scattering contrast in biological media such as skin, brain, and fatty tissues tend to not exhibit significant variation with wavelength.[6] This hampers the capabilities of label-free scattering-based microscopes in biological specimens. For this reason, many applications require ex-vivo sample preparation coupled with exogenous dyes to provide chromophore-specific visualizations in biological samples. A common example of this is hematoxylin and eosin (H&E) staining of tissues frequently used during histological assessment.[7] Unfortunately, generating stained preparations may be undesirable as it can be time consuming and may alter biochemistry and biological structures. For example, when preparing tissues for H&E staining, lipids structures are removed completely.[7]

Absorption based modalities such as photoacoustic microscopy (PAM),[8] fluorescence,[9,10] and multiphoton fluorescence,[10] leverage absorption interactions to provide visualizations of chromophores. Absorption contrast in biological tissues tends to be highly chromophore specific, where most molecules present unique absorption spectra.[6,11] Therefore, optical absorption microscopy is particularly attractive for label-free imaging of biological samples, where endogenous absorption profiles can be leveraged to provide information on properties such as chemical bonding,[12] sample composition,[13] and temperature.[14]

Absorbed energy may be dissipated by chromophores through either optical radiation (radiative) or non-radiative relaxation, most absorption imaging mechanisms may be broadly classified into these corresponding subcategories. During non-radiative relaxation, absorbed optical energy is converted into heat. If the excitation event is sufficiently rapid, this heating may cause thermoelastic expansion resulting in localized photoacoustic pressures.[15] These temperature rises are leveraged in photothermal modalities,[16] while the pressure rises are leveraged in photoacoustic imaging.[8,11,15] In traditional PAM, pressure waves are allowed to propagate through the sample as ultrasound waves, which are then detected at the sample surface with ultrasound transducers.[8,11,15] PAM has demonstrated label-free visualizations of a wide range of endogenous chromophores, including DNA,[17,18] lipids,[19–21] and hemeproteins.[22,23] During radiative relaxation, absorbed optical energy is released through the emission of photons. Generally, emitted photons exhibit a different energy level compared to the absorbed photons. Radiative relaxation contrast encompasses a variety of mechanisms such as stimulated Raman scattering,[24] fluorescence,[9,10] and multiphoton fluorescence.[10] For example, in multiphoton fluorescence imaging, the energy of two or more photons is absorbed then released as a higher energy fluorescent photon. In practice, a range of biomolecules including NADPH, collagen, and elastin, have been visualized label-free with such radiative absorption techniques.[25]

To further provide label-free visualizations of biomolecules in complex media it would be desirable to have a technique which could capitalize on the advantages of scattering contrast and both radiative and non-radiative absorption modalities. Ideally, capturing all contrasts simultaneously. Here, a second-generation of Photoacoustic Remote Sensing (PARS) microscopy is presented, entitled total-absorption PARS (TA-PARS), which facilitates label-free non-contact capture of scattering, radiative absorption, and non-radiative absorption simultaneously. Unlike traditional radiative or non-radiative absorption modalities where contrast may be dictated by efficiency factors such as the photothermal conversion efficiency or fluorescence quantum yield, the TA-PARS may capture nearly all the optical properties of a chromophore, providing simultaneous sensitivity to most chromophores. By extension, capturing both radiative and non-radiative absorption fractions may also yield additional information. The ratio of the two absorption fractions is expected to provide an additional chromophore specific metric. This ratio of the radiative to non-radiative absorption fractions is proposed as the quantum efficiency ratio (QER). In biomolecules such as collagen, and DNA, the QER may enhance chromophore specific recovery. To the best of our knowledge, this array of imaging contrast (optical scattering, radiative absorption, non-radiative absorption) has not yet been provided by any other independent imaging modality (see more information in Supplementary Information Section 1: Table 1).

In TA-PARS a picosecond scale pulsed excitation laser elicits radiative and nonradiative (thermal and pressure) perturbations in a sample. The thermal and pressure perturbations generate corresponding modulations in the local optical properties. A secondary probe beam co-focused with the excitation captures the non-radiative absorption induced modulations to the local optical properties as changes in backscattering intensity (Fig. 1).[23,26] These backscatter modulations are then directly correlated to the local non-radiative absorption contrast.[23,26] By the nature of the probe architecture, the unperturbed backscatter (pre-excitation event) also captures the scattering contrast as seen by the probe beam (Fig. 1).[27] Unlike traditional photoacoustic methods, rather than relying on the pressure waves to propagate through the sample before detection via acoustic transducer, the TA-PARS probe may instantaneously detect the induced modulations at the excited location.[23,26] Therefore, TA-PARS offers non-contact operation, facilitating imaging of delicate, and sensitive samples, which would otherwise be impractical to image with traditional contact-based PAM methods. Since TA-PARS, like PAM, relies only on the generation of heat and subsequently pressure to provide contrast, the absorption mechanism is non-specific, and highly sensitive to small changes in relative absorption.[23,26] This allows any variety of absorption mechanisms such as vibrational absorption,[12] stimulated Raman absorption,[28] and electronic absorption,[15] to be detected with PARS and PAM. Previously, PARS has demonstrated label-free non-radiative absorption contrast of hemoglobin, DNA, RNA,[18,29–33] lipids,[20,21] and cytochromes,[29,30] in specimens such as chicken embryo models,[20] resected tissue specimens,[29] and live murine models.[23,34,35] In TA-PARS, a unique secondary detection pathway captures radiative relaxation contrast, in addition to the non-radiative absorption. The radiative absorption pathway was designed to broadly collect all optical emissions at any wavelength of light, excluding the excitation and detection. As a result, the radiative detection pathway captures non-specific optical emissions from the sample. Radiative relaxation signals may then be attributed to any number of radiative effects such as spontaneous Raman scattering, stimulated Raman scattering, autofluorescence, multiphoton autofluorescence, etc. The potential interactions captured during a TA-PARS excitation event are outlined in Fig. 1.

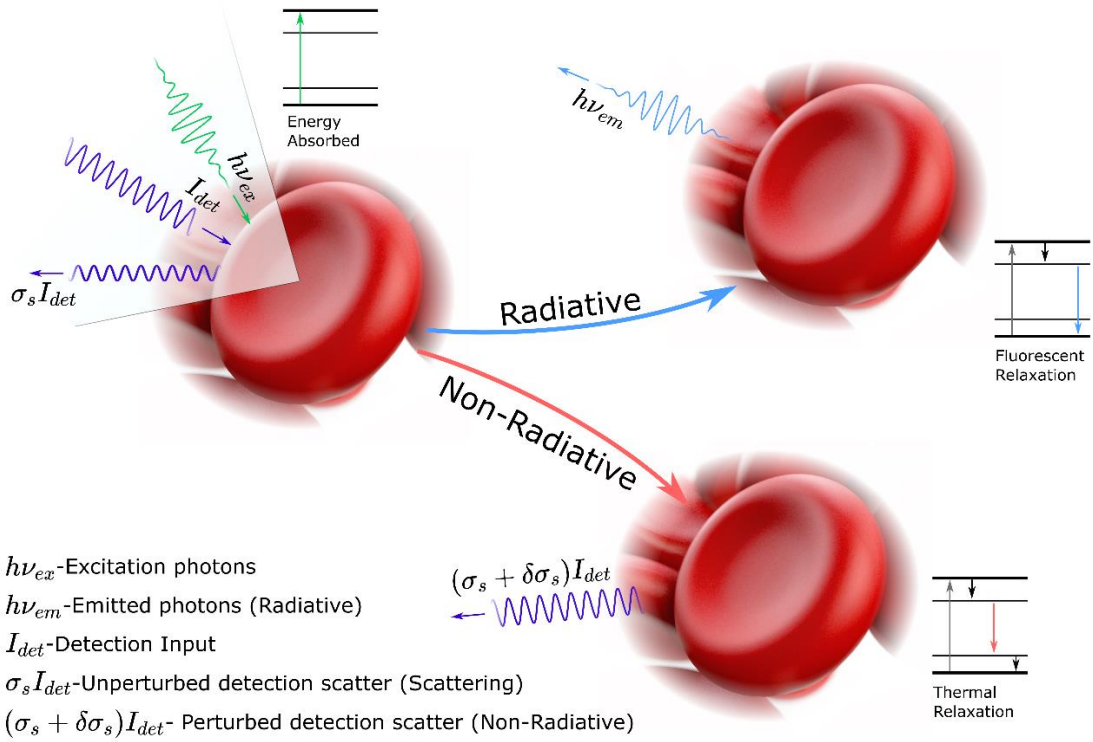

**Fig 1.** TA-PARS contrast mechanisms. Each excitation will always generate some fraction of radiative and non-radiative relaxation effects. The non-radiative relaxation leads to heat and pressure induced modulations, which in turn cause back-reflected intensity variations in the detection beam. PARS signals are denoted as some change in reflectivity multiplied by the incident detection ($RI_{det}$). The radiative absorption pathway captures optical emissions attributed to radiative relaxation such as stimulated Raman scattering, fluorescence, multiphoton fluorescence, etc. Emissions are denoted as some wavelength and energy optical emission ($h\nu_{em}$). The local scattering contrast is captured as the unmodulated backscatter (pre-excitation pulse) of the detection beam. The scattering contrast is denoted as the unperturbed scattering profile multiplied by the incident detection power ($\sigma_s I_{det}$).

Previously, multimodal fluorescence microscopy and PAM have leveraged similar absorption mechanisms, capturing both radiative fluorescence contrast and non-radiative photoacoustic contrast.[36–40] Such multimodal PAM and fluorescence microscopy systems have been inherently complicated, due to the acoustic detection mechanism of traditional PAM. In many cases, it may be challenging to guide the excitation pulses and emitted radiative relaxation light through or around the acoustic transducer. The need for acoustic coupling may also render these multimodal systems impractical for imaging delicate, and sensitive samples. Recently, Zhou et al. showed a non-contact dual-modality PARS and dye-based fluorescence microscope.[41] The dual-modal system used Rhodamine B dye to generate fluorescence contrast to complement the PARS visualizations. In contrast, the TA-PARS captures simultaneous complementary radiative and non-radiative contrast label-free from a range of endogenous chromophores.

To improve the sensitivity of the TA-PARS and facilitate the detection of radiative absorption contrast, a variety of systematic changes are introduced compared to previously reported PARS systems. The presented TA-PARS features 266 nm and 515 nm excitation, providing sensitivity to DNA, heme proteins, NADPH, collagen, elastin, amino acids, and a variety of fluorescent dyes. The TA-PARS features a specific optical pathway with dichroic filters and avalanche photodiode, to isolate and detect the radiative absorption contrast. The TA-PARS probe beam was implemented with a 405 nm laser diode. This probe wavelength provides improved scattering resolution, which improves the confocal overlap between the PARS excitation and detection spots on the sample. Combined with a circulator-based probe beam pathway and avalanche photodetector, the TA-PARS provides improved sensitivity compared to previous implementations. The visible wavelength probe also provides improved compatibility between the visible and UV excitation wavelengths. The prevalence of chromatic aberrations was suppressed when using achromatic refractive optics by reducing the disparity in the excitation and detection wavelengths, as opposed to previous comparable NIR based PARS systems.[18,29–34,42]

The TA-PARS imaging contrast was explored in simple dye samples, unprocessed resected tissue specimens, and sections of preserved human tissues. The TA-PARS captures label-free features such as adipocytes, fibrin, connective tissues, neuron structures, and cell nuclei. Visualizations of intranuclear structures are captured with sufficient clarity and contrast to identify individual atypical nuclei. One potential proposed application of TA-PARS, label-free histological imaging, was explored in unstained sections of human tissues. TA-PARS visualization fidelity is assessed through one-to-one comparison against traditional H&E-stained images. The TA-PARS total-absorption and QER contrast mechanisms are also validated in a series of dye and tissue samples. Results show high correlation between radiative relaxation characteristics and TA-PARS-measured QER in a variety of fluorescent dyes, and tissues. These QER visualizations are used to extract regions of specific biomolecules such as collagen, elastin, and nuclei in tissue samples. This enables realization of a broadly applicable high resolution absorption contrast microscope system. The TA-PARS may provide unprecedented label-free contrast in any variety of biological specimens, providing otherwise inaccessible visualizations.

## Methods

*System Architecture*

The architecture of the experimental system is illustrated in Fig. 2. Two excitation lasers were employed in the proposed architecture. The first was a 515 nm visible excitation (Fig. 1(i)), the second was a 266 nm UV excitation (Fig. 2(ii)). The 515 nm excitation uses a 50 kHz to 2.7 MHz 2 ps pulsed 1030 nm fiber laser (YLPP-1-150-v-30, IPG Photonics). The second harmonic was generated with a lithium triborate crystal. The 515 nm harmonic was separated via dichroic mirror, then spatial filtered with a pinhole prior to use in the imaging system. The 266 nm excitation uses a 50 kHz 400 ps pulsed diode laser (Wedge XF 266, RPMC). Output from the 266 nm laser was separated from residual 532 nm excitation using a prism, then expanded prior to use in the imaging system.

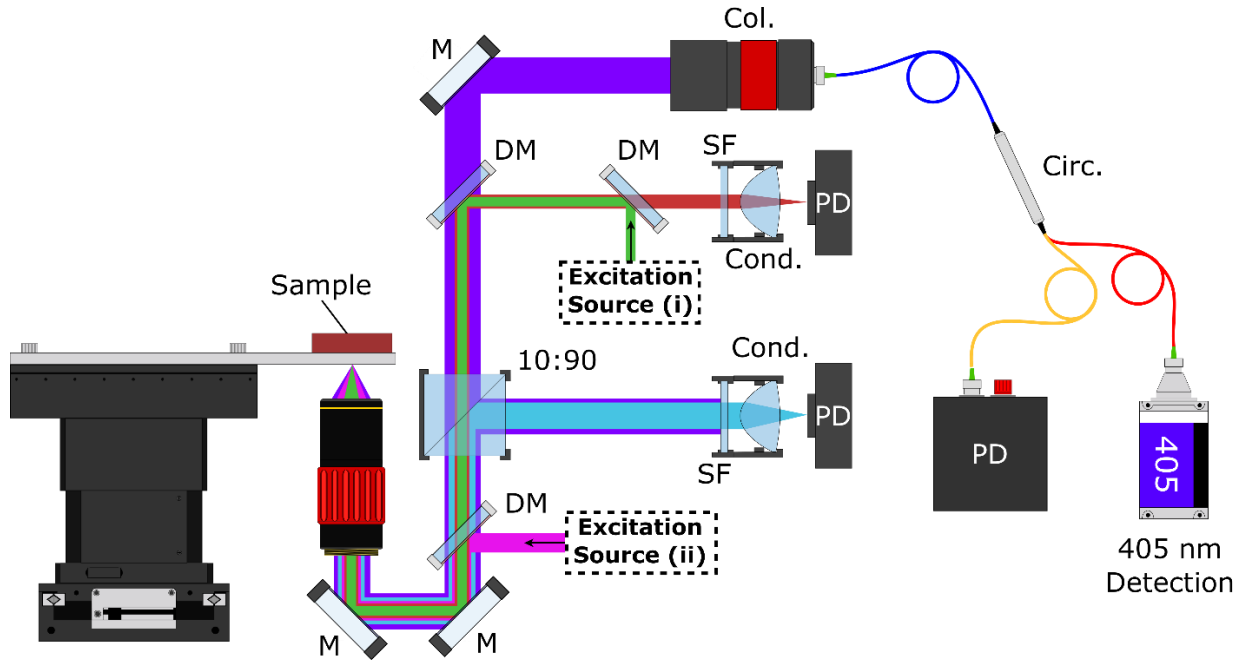

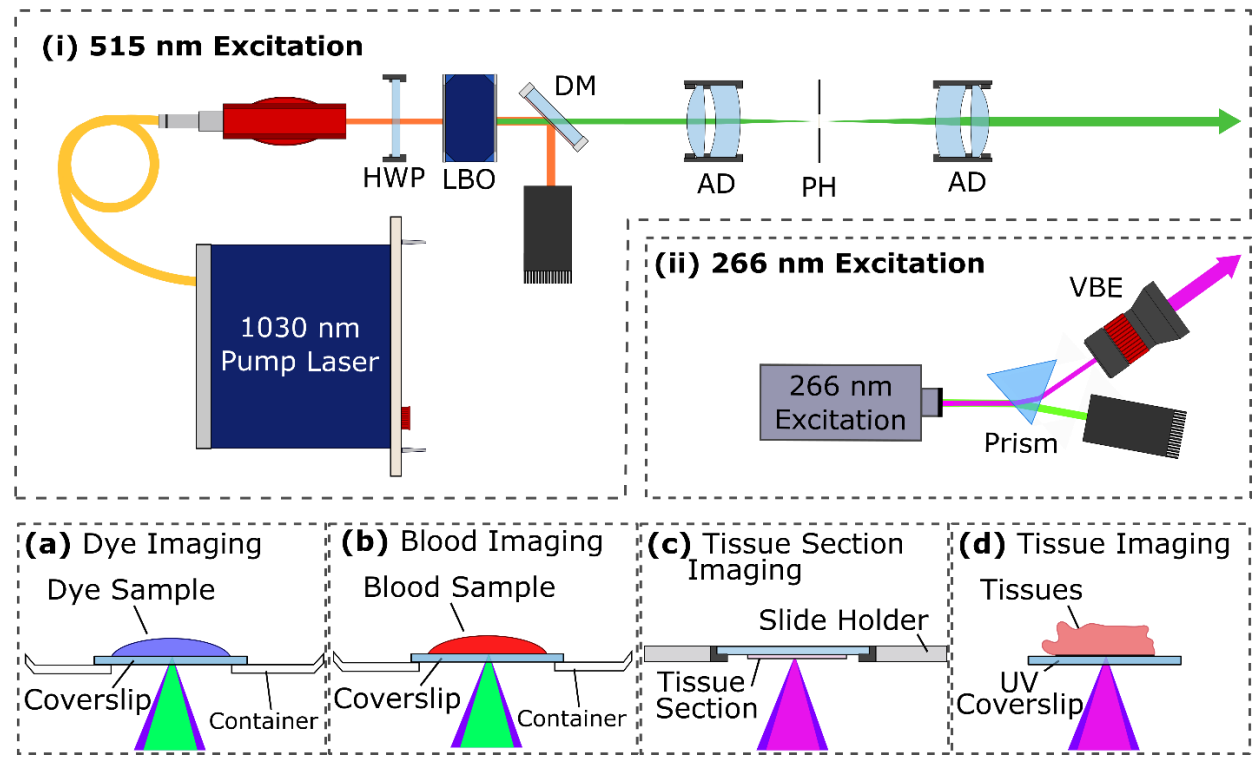

**Fig. 2** Simplified experimental system setups (i) 515 nm excitation system (b) 266 nm excitation system. (a-d) Sample configuration for imaging each type of tissue. Component labels are as follows: half wave plate (HWP), lithium triborate crystal (LBO), dichroic mirror (DM), air spaced doublet (AD), pinhole (PH), variable beam expander (VBE), collimator (Col.), circulator (Circ.), spectral filter (SF), condenser lens (Cond.), photodiode (PD), 10:90 splitter (90:10), mirror (M).

The two excitations share a common 405 nm PARS detection system. The 405 nm detection light was provided by a 405 nm OBIS-LS laser (OBIS LS 405, Coherent). The detection was fiber coupled through a circulator into the system, where it was combined with the excitations via dichroic mirror. The combined excitation and detection are co-focused onto the sample using a 0.42 NA UV objective lens. Sample configurations are outlined in Fig. 2a to Fig. 2d. Back-reflected detection from the sample returns to the circulator by the same path as forward propagation. The back-reflected detection contains the PARS non-radiative absorption contrast as nanosecond scale intensity modulations which are captured with a photodiode.

Radiative relaxation from each of the 266 nm and 515 nm excitation were independently captured with different photodiodes. The 515 nm induced radiative relaxation was isolated with dichroic mirrors, then captured using a photodiode. The 266 nm induced radiative relaxation was isolated by redirecting 10% of the total light intensity returned from the sample towards a photodetector. This light was then spectrally filtered to remove residual excitation and detection prior to measurement.

*Image Formation*

To form an image, the mechanical stages were used to scan the sample over the objective lens. The excitation sources were continuously pulsed at 50 kHz, while the stage velocity was regulated to achieve the desired pixel size (spacing between interrogation events). Each time the excitation laser was pulsed, a collection event was triggered. During a collection event, a few hundred nanosecond segment was collected from 4 input signals using a high-speed digitizer (RZE-004-200, Gage Applied). These signals were, the PARS scattering signal, the PARS non-radiative relaxation signal, the PARS radiative relaxation signal, and a positional signal from the stages. The time resolved scattering, absorption, and position signals, were then compressed down to single characteristic features. This serves to substantially reduce the volume of data capture during a collection. To reconstruct the absorption and scattering images, the raw data was fitted to a Cartesian grid based on the location signal at each interrogation. Raw images were then gaussian filtered, and rescaled based on histogram distribution prior to visualization.

*Sample Preparation*

This study features several types of human and rodent tissues, including formalin fixed paraffin embedded tissues, formalin resected tissues, and unprocessed resected tissues. Human tissues were obtained by Clinical collaborators at the Cross-Cancer Institute (Edmonton, Alberta, Canada), tissues were collected from anonymous patient donors with all patient identifiers removed from the samples. The ethics committees waived the requirement for patient consent on the condition that samples were archival tissue no longer required for diagnostic purposes, and that no patient identifiers were provided to the researchers. The samples were obtained under a protocol approved by Research Ethics Board of Alberta (Protocol ID: HREBA.CC-18-0277) and University of Waterloo Health Research Ethics Committee (Photoacoustic Remote Sensing (PARS) Microscopy of Surgical Resection, Needle Biopsy, and Pathology Specimens; Protocol ID: 40275). All human tissue experiments were performed in accordance with the relevant guidelines and regulations. Murine and Rattus tissues were obtained from Charles River Brown Norway rats, and Charles River SKH-1 mice under a protocol approved by the University of Waterloo Health Research Ethics Committee (Photoacoustic Remote Sensing (PARS) Microscopy of Resected Rodent Tissues; Protocol ID: 41543). All murine tissue experiments were performed in accordance with the relevant guidelines and regulations. The preparation methods for all sample types are described below.

Formalin Fixed Paraffin Embedded (FFPE) thin tissue sections preparation

Bulk resected tissues were submerged in formalin fixative solution within 20 minutes of excision. Tissues were fixed for 24 to 48 hours. Once fixed, tissues were dehydrated using increasing concentrations of ethanol. To permit paraffin wax penetration, tissues were then cleared with xylene to remove ethanol and residual fats. Tissues were then embedded into paraffin wax forming FFPE tissue blocks. Several adjacent thin sections of tissue (~5 µm thick) were sliced from the surface of the FFPE tissue block using a microtome. The thin sections were placed onto glass slides and baked at 60 °C to remove excess paraffin. The unstained thin sections were transported to the PhotoMedicine Labs at the University of Waterloo where they underwent PARS imaging. Once imaged, specimens were stained with H&E dyes, and covered with mounting media and a coverslip. Stained slides were imaged using a transmission mode brightfield microscope. This provided a direct comparison between PARS imaging and the gold standard of H&E staining.

Formalin fixed resected tissue sample preparation

Bulk resected tissues were obtained with the aid of collaborators at the Central Animal Facility, University of Waterloo under animal care protocol (Photoacoustic Remote Sensing (PARS) Microscopy of Resected Rodent Tissues; Protocol ID: 41543). Resected tissues were submerged in formalin fixative solution within 20 minutes of excision. Tissues were imaged across a range of times from ~20 minutes to ~3 hours of excision.

Dye samples

A drop of dye was placed directly onto a transparent imaging window inside of an enclosed container. The container was covered to prevent drying and evaporation of the dye samples. PARS signals were then captured through the UV coverslip across a one-millimeter squared area.

**Results and Discussion**

*Total-Absorption PARS System Improvements*

Several improvements have been implemented over previous PARS embodiments. The TA-PARS features a novel 405 nm detection source, as opposed to the near-infrared (750 nm to 1500 nm) detection sources featured in previous PARS embodiments.[18,29–34,42] Compared to the previous infrared detections, the diffraction limited focus of the 405 nm improves the confocal overlap between the excitation and detection focal spots on the sample.[20,43] Improving the confocal overlap may contribute to higher imaging sensitivity, by exciting a larger fraction of the detection focal spot, thereby improving PARS modulation efficiency.[20,43] Additionally, the 405 nm backscattering in biological tissues tends to be significantly stronger than that of near-infrared wavelengths.[6] In conjunction with these mechanism improvements, the avalanche photodiode used in the TA-PARS provides an order of magnitude more responsivity compared to previous photodetectors.[18,29–34,42] These refinements culminate in a significant reduction in the excitation and detection energies required for imaging. The TA-PARS results shown herein, were captured with detection powers as low as 156 µW, and excitation pulse energies as low as 400 pJ. These values are ~15x and ~2x lower, respectively, than the lowest excitation and detection powers previously reported by any other PARS system (see more information in Supplemental Information Section 2: Table 2). Resolution was captured as ~350 nm mean resolution, from the edge spread function (10% to 90% rise) generated from imaging FFPE nuclei, and the line spread function generated from imaging myelin fibers (FWHM) (see more information in Supplemental Information Section 3: Fig. 2). As well, a qualitative comparison of the TA-PARS with previous PARS embodiments can be found in the Supplemental Information Section 2: Fig. 1. The improvements in image clarity and consistency lend to the efficacy of the TA-PARS over previously reported embodiments.

*Imaging of thin tissue sections*

The performance of the 266 nm excitation TA-PARS was first characterized in thin sections of formalin fixed paraffin embedded (FFPE) human brain tissues. The TA-PARS non-radiative relaxation visualization is shown in Fig. 3(a), while the radiative relaxation is shown in Fig. 3(b). The non-radiative relaxation signals were captured based on nanosecond scale pressure- and temperature-induced modulations in the collected backscattered 405 nm detection beam from the sample. The radiative absorption contrast was captured as optical emissions from the sample, excluding the excitation and detection wavelengths which were blocked by optical filters. Concurrently, the unperturbed backscatter of the 405 nm probe captures the local optical scattering from the sample (Fig. 3 (c)) (full sized images available in Supplementary Information Section 4: Fig. 3). With this contrast, most of the salient tissue structures were captured. The non-radiative absorption contrast highlights predominately nuclear structures, while the radiative contrast captures extranuclear features. The optical scattering contrast captures the morphology of the thin tissue section. In resected tissues this scattering contrast becomes less applicable, and hence was not explored in other samples.

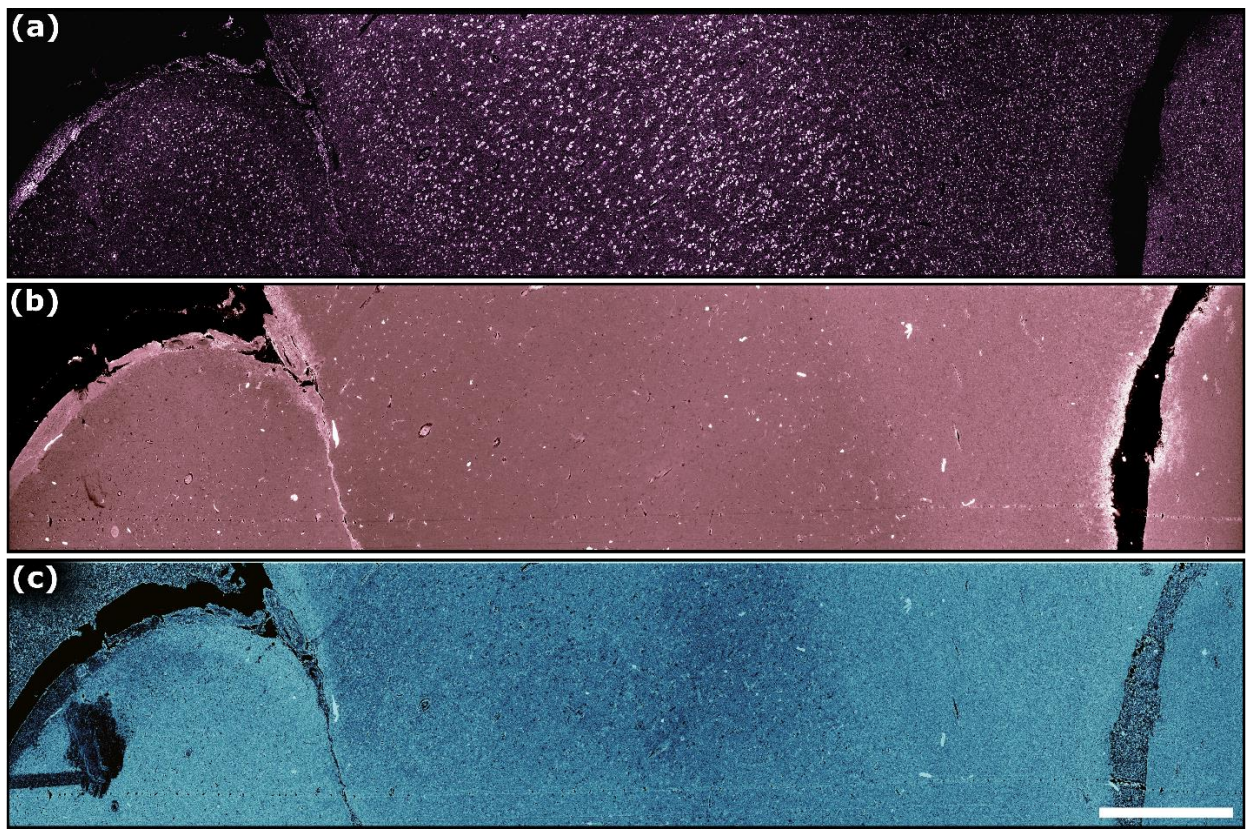

**Fig. 3** Comparison of the three different contrasts (non-radiative absorption, radiative absorption, and scattering) provided by the TA-PARS system, in a thin section of preserved human brain tissues. (a) 266 nm non-radiative absorption contrast. (b) 266 nm radiative absorption contrast. (c) 405 nm scattering contrast. Scale Bar: 1 mm.

*Imaging in bulk resected tissues*

Next, the TA-PARS was explored on a variety of unprocessed tissue preparations. In resected human skin tissues, the TA-PARS captures the epithelial layer at the margin of the resected tissues (Fig. 4). In the upper left, the sloughing layers of dead skin tissues were captured in the radiative and non-radiative visualizations concurrently (Fig. 4(a)). The radiative visualization appears to provide improved contrast in recovering these tissue layers as compared to the non-radiative image. In another subcutaneous region of the resected human skin tissues, the TA-PARS captures connective tissues, with sparse nuclei, and elongated fibrin features (Fig. 4(b)). Traditionally these connective tissue structures are not clearly differentiable in H&E preparations since there is a high concentration of lipids which are removed during paraffin preparation. Of particular interest is the presence of fibrin. Fibrin is indicative of wound healing which likely began directly following tissue resection. Fibrin structures are traditionally difficult to recover in H&E images.[44] However, these appear to be recoverable when directly imaging the resected tissues with the TA-PARS.

The proposed system was also applied to imaging resected unprocessed rattus brain tissues. The TA-PARS acquisition highlights the granular layer in the brain revealing dense regions of nuclear structures (Fig. 4(c)). The nuclei of the granular layer are presented with higher contrast relative to surrounding tissues in the non-radiative image as compared to the radiative representation. Since nuclei do not provide significant radiative contrast,[45] the nuclear structures in the radiative image appear as voids or lack of signal within the specimen. While some potential nuclei may be observed, they may not be identified with significant confidence, as compared those in the TA-PARS non-radiative representation. Along the top right of the non-radiative acquisition, structures resembling myelinated neurons can be identified surrounding the more sparsely populated nuclei in that area. Further acquisitions in neighboring regions accentuate the apparent myelinated neuron structures (Fig. 4(d)). Dense structures indicative of the web of overlapping and interconnected neurons are apparent within these regions, where tightly woven neurons are observed arranged around a void in the tissue. Then, zooming out to a larger nearby imaging field, sections of distinct tissues were recovered with the non-radiative contrast (Fig. 4(e)). The left side of the field contains dense bundles of myelin with more sparse nuclei, as opposed to the right side which contains more sparse myelinated structures with increased nuclear density (full sized images from Fig. 4 are available in Supplementary Information Section 4: Fig. 4 to Fig. 6).

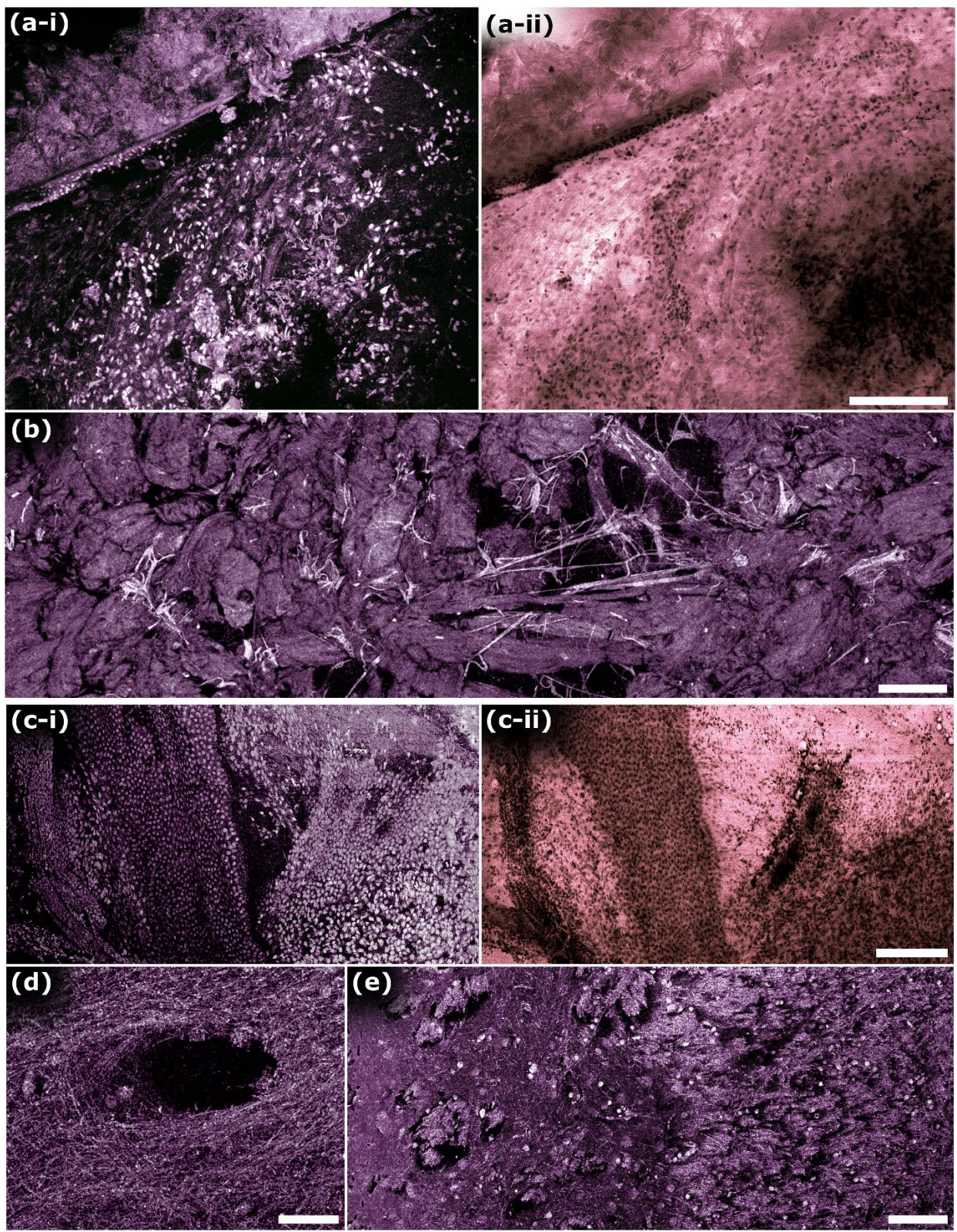

**Fig. 4** Total-absorption second-generation photoacoustic remote sensing (TA-PARS) microscopy of unstained and unprocessed resected tissue specimens. (a) Resected human skin tissues. (a-i) TA-PARS non-radiative absorption contrast. (a-ii) TA-PARS radiative contrast of the same region of tissues. Scale Bar: 200 μm (b) Deep subcutaneous

connective tissues of resected human skin tissues with elongated strings of fibrin. Scale Bar: 100 µm. (c) TA-PARS image of unprocessed resected Rattus brain tissues (c-i) TA-PARS non-radiative absorption contrast. (c-ii) TA-PARS radiative contrast image of the same region. Scale Bar: 200 µm (d) TA-PARS non-radiative absorption contrast image highlighting apparent myelinated neuron structures within the brain tissues. Scale Bar: 50 µm (e) TA-PARS non-radiative absorption image of the boundary between two regions of brain tissues. Scale Bar: 100 µm.

*Potential TA-PARS application: Label-Free Histology Imaging*

The unique contrast provided by the TA-PARS system may be envisioned for use in several applications. One such application, label-free histopathology of tissues, was explored here. In a clinical context, the TA-PARS mechanism provides an opportunity to accurately emulate traditional histochemical staining contrast, specifically H&E staining. The non-radiative TA-PARS signal contrast is analogous to that provided by the hematoxylin staining of the cell nuclei (Fig. 5(a)). Here, a section of FFPE human brain tissue was imaged with the non-radiative PARS (Fig. 5(a-i)). This non-radiative information was then colored to emulate the contrast of hematoxylin staining (Fig. 5(a-ii)). The same tissue section was then stained only with hematoxylin and imaged under a brightfield microscope (Fig. 5(a-iii)), providing a direct one-to-one comparison. These visualizations are expected to be highly similar since the primary target of hematoxylin stain and the non-radiative portion of TA-PARS is nuclei, though other chromophores will also contribute to some degree. A similar approach was applied to eosin staining in an adjacent section. The adjacent section was imaged with the radiative PARS (Fig. 5(b-i)). This radiative information was then colored to emulate the contrast of eosin staining (Fig. 5(b-ii)). This section was then stained with eosin (Fig. 5(b-iii)), providing a direct one-to-one comparison of the radiative contrast and eosin staining. In each of the TA-PARS and eosin-stained images, analogous microvasculature and red blood cells were resolved throughout the brain tissues. These visualizations are expected since the primary targets of the radiative portion of TA-PARS include hemeproteins, NADPH, flavins, collagen, elastin, and extracellular matrix, closely mirroring the chromophores targeted by eosin staining of extranuclear materials.

As the different contrast mechanisms of the TA-PARS closely emulate the visualizations of H&E staining, the proposed system may provide true H&E-like contrast in a single acquisition. This technique is substantially faster than the previously reported dual-excitation PARS system proposed by Bell et al.[29] and Ecclestone et al.[30] Moreover, the TA-PARS may provide substantially improved visualizations compared to previous PARS emulated H&E systems which relied on scattering microscopy to estimate eosin-like contrast.[27,46] The scattering microscopy-based methods are unable to provide clear images in complex scattering samples such as bulk resected human tissues. In contrast, the TA-PARS can directly measure the extranuclear chromophores via radiative contrast mechanisms,[25] thus providing analogous contrast to H&E regardless of specimen morphology. Here, the different TA-PARS visualizations were combined using a linear color mixture to generate an effective representation of traditional H&E staining within unstained tissues. An example in resected FFPE human brain tissue is shown in Fig. 5c. The wide field image highlights the boundary of cancerous and healthy brain tissues. To qualitatively compare the TA-PARS to traditional H&E images, a series of human breast tissue sections was scanned with the TA-PARS (Fig. 5(d-i) and Fig. 5(e-i)), then stained with H&E dyes and imaged under a brightfield microscope (Fig. 5(d-ii) and Fig. 5(e-ii)). The TA-PARS emulated H&E visualizations are effectively identical to the H&E preparations. In both images, clinically relevant features of the metastatic breast lymph node tissues are equally accessible (full sized images from Fig. 5 are available in Supplementary Information Section 4: Fig. 7 to Fig. 12).

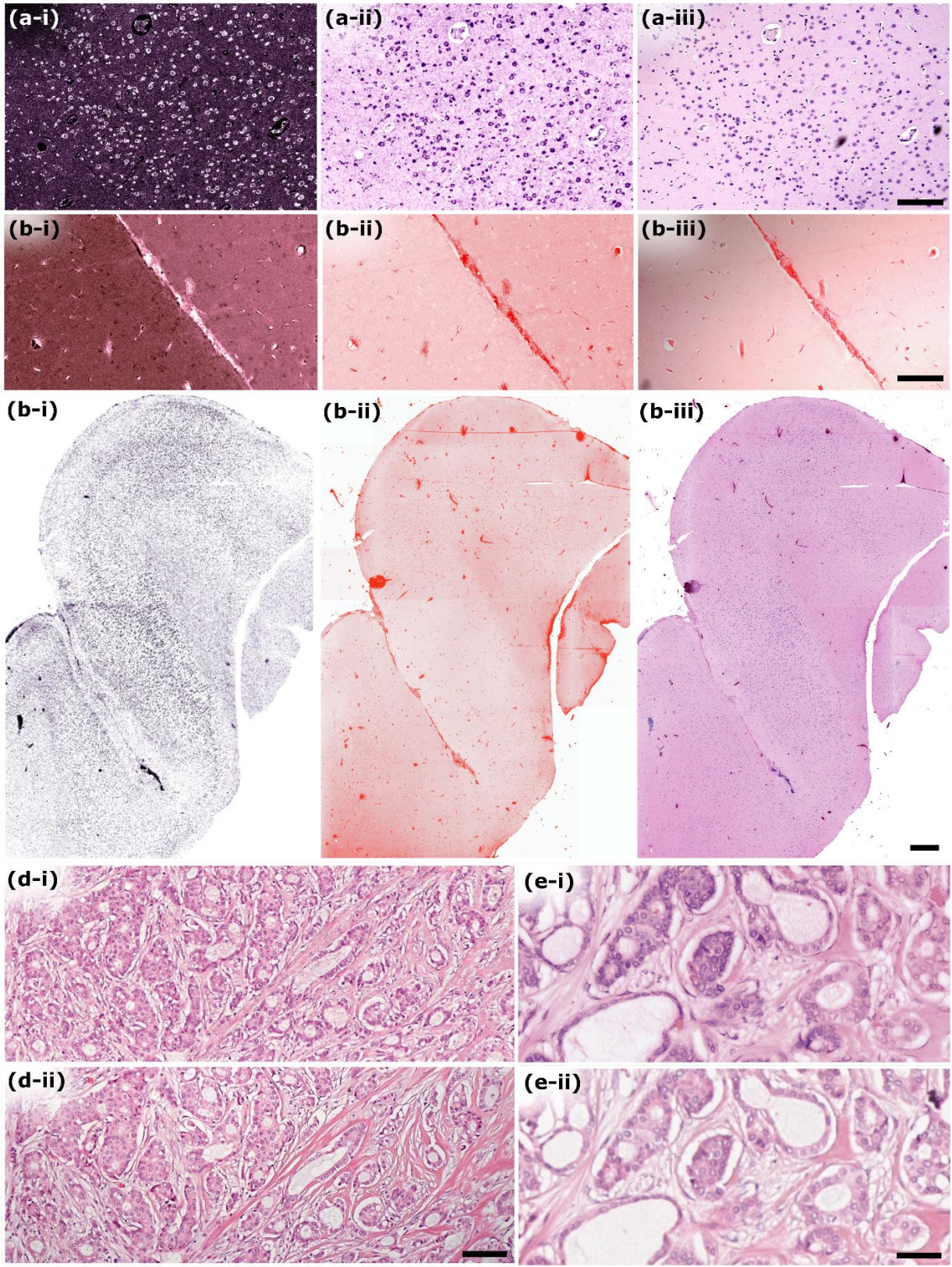

**Fig. 5** Comparison of the SG-PARS radiative and non-radiative absorption contrast with traditional histochemical stains. (a) Comparison of TA-PARS 266 nm non-radiative absorption contrast with hematoxylin staining. (a-i) TA-

PARS 266 nm non-radiative absorption contrast highlighting predominately nuclear structures. (a-ii) False colored version of image presented in (a-i). (a-iii) Same section of tissue stained with hematoxylin stain only providing a one-to-one comparison with non-radiative TA-PARS. Scale Bar: 200 µm. (b) Comparison of TA-PARS 266 nm radiative absorption contrast with eosin staining. (b-i) TA-PARS 266 nm radiative absorption contrast highlighting predominately extra-nuclear structures (i.e. collagen, elastin, NADPH). (b-ii) False colored version of image presented in (b-i). (b-iii) Same section of tissue stained with eosin stain only providing a one-to-one comparison with radiative TA-PARS. Scale Bar: 200 µm. (c) TA-PARS image of nearly an entire section of resected brain tissues. (c-i) Non-radiative absorption contrast of predominately nuclear structures. (c-ii) Radiative absorption contrast of predominately extra-nuclear structures. (c-iii) TA-PARS emulated H&E image of nearly the entire section of resected brain tissues with analogous contrast to traditional H&E staining. Scale Bar: 1 mm. (d & e) One-to-one comparison between standard brightfield H&E preparations (i) and TA-PARS emulated H&E images (ii) in thin sections of human breast tissues. (d) Scale Bar: 100 µm. (e) Scale Bar: 50 µm

*PARS QER Imaging*

Finally, the proposed QER imaging mode was characterized in a series of dyes and tissues. The QER or the ratio of the non-radiative and radiative absorption fractions is expected to contain further biomolecule-specific information. Ideally, the local absorption fraction should correlate directly with radiative relaxation properties. Here, the TA-PARS was applied to measure a series of fluorescent dyes with varying quantum efficiencies. The 515 nm excitation was used to generate radiative and non-radiative relaxation signals which were captured simultaneously. The relative radiative and non-radiative signal intensities are then plotted in Fig. 6(a-i). The QER (quotient of the relative radiative signal intensity over the relative non-radiative signal intensity) is then plotted against reported quantum efficiency (QE) values for the samples (Fig. 6(a-ii)). The radiative PARS signals ($P_r$) are expected to increase linearly with the QE ($P_r \propto QE$), while the non-radiative PARS ($P_{nr}$) signals are expected to decrease linearly with QE ($P_{nr} \propto 1 - QE$). Therefore, the fractional relationship between the non-radiative and radiative signals is represented by the quotient of the linear functions ($QER = P_{nr}/P_r \propto QE/(1 - QE)$). The empirical results fit well to this expected model ($R = 0.988$).

The QER acquisition process was then applied to imaging of thin sections of FFPE human tissues. Based on the non-radiative (Fig. 6(b-i)) and radiative (Fig. 6(b-ii)) signals, the QER was calculated for each image pixel, generating a QER image (Fig. 6(b-iii)). The result represents a dataset encoding chromophore-specific attributes, in addition to the independent absorption fractions. The QER processing helps to further separate otherwise similar tissue types from solely the radiative or non-radiative acquisitions. A colorized version of the QER image is shown in Fig. 6(c-i) to highlight various tissue components. The high QER biomolecules (DNA, RNA, etc.) appear as a light blue color, while the low QER biomolecules (collagen, elastin, etc.) appear pink, and purple. Compared to the H&E visualization captured following the QER imaging session (Fig. 6(c-ii)), collagen and elastin (dark red) composing the fibrous connective tissues are easy to identify due to their low QER. Conversely, nuclear structures are appreciable in blue due to their high QER. The intralobular connective tissues surrounding the nuclei are also differentiated from the fibrous connective tissues in purple in the QER visualization as compared to the H&E-stained image (full sized images available in Supplementary Information section 4: Fig. 12 and Fig. 13). In calculating the QER from the TA-PARS a complementary imaging contrast is provided, enabling further chromophore specificity than is accessible with radiative or non-radiative modalities independently.

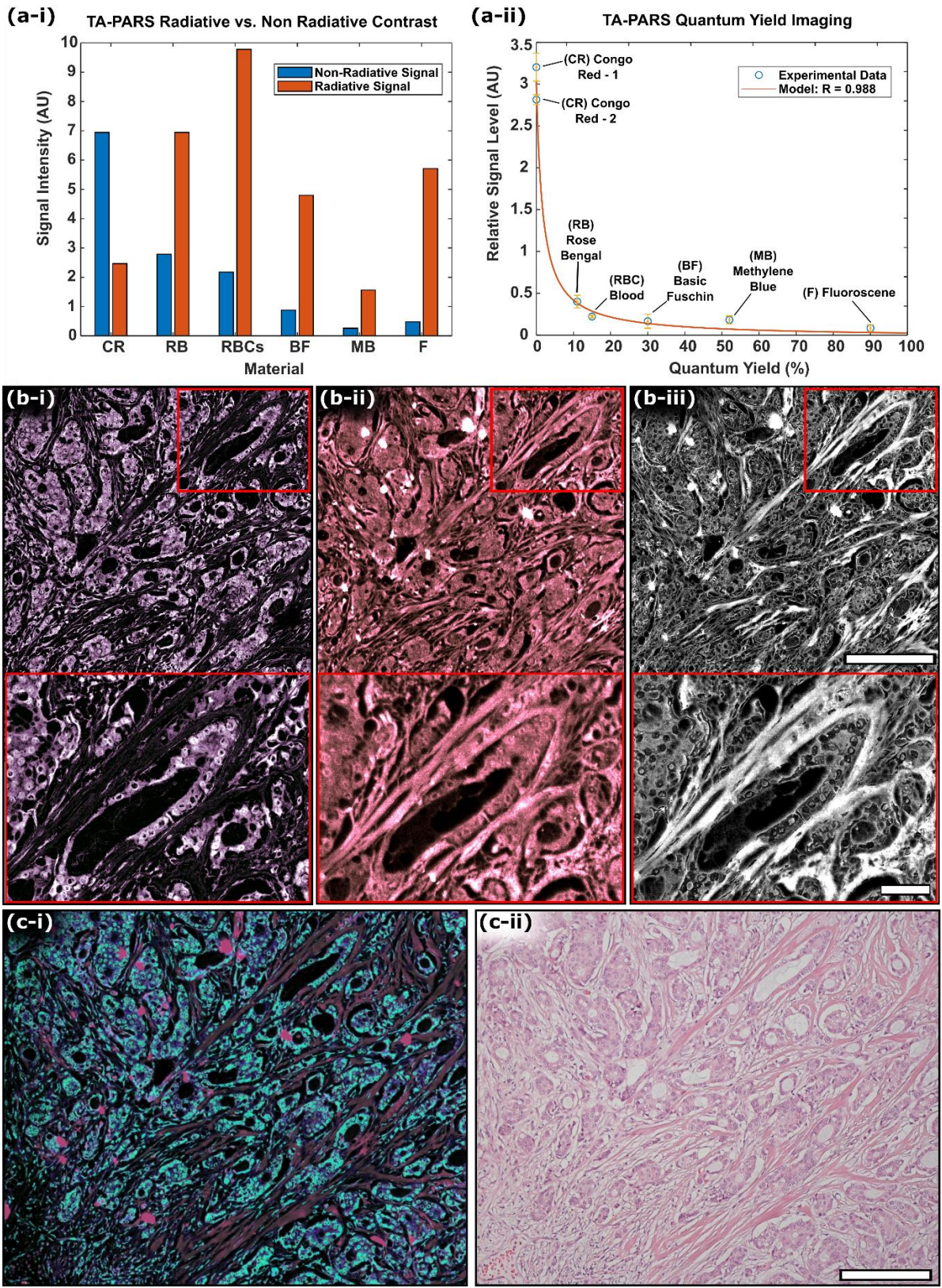

**Fig. 6** TA-PARS quantum efficiency ratio (QER) measurement results. (a-i) PARS radiative vs non-radiative signal magnitude in a series of samples with varying Quantum Yields. (a-ii) Measured QER from the TA-PARS signals,

compared to the expected relationship. (b) An example of the TA-PARS QER imaging in resected tissue specimens. (b-i) TA-PARS non-radiative absorption contrast image. (b-ii) TA-PARS radiative absorption contrast image. (b-iii) TA-PARS quantum yield image. Scale Bar (Upper): 200 µm. Scale Bar (Lower): 50 µm (c-i) Artificially color mapped image of tissues where the quantum yield is used to identify different tissue structures. (c-ii) H&E image of the same section of tissues for comparison of tissue structures. Scale Bar: 200 µm.

*Further Considerations*

While this work focuses primarily on label-free TA-PARS contrast, this technique may also provide sensitivity to a broad range of current stains. Some biomolecules may not exhibit sufficiently unique endogenous absorption or scattering properties to differentiate them against other constituents. Here the TA-PARS may leverage a broad selection of chromophore-specific exogenous contrast agents previously developed for other absorption techniques, such as fluorescent or photoacoustic dyes. As exemplified here, the TA-PARS may provide contrast in a range of fluorescent dyes with highly varied QE properties. This effect may be leveraged to differentiate between two dyes of similar emission spectra but differing quantum yields.

## **Conclusions**

In summary, a second-generation PARS microscope was presented. Entitled TA-PARS, the system captures absorption and scattering effects in a variety of samples. With the suite of contrast (optical scattering, radiative absorption, and non-radiative absorption) provided by the TA-PARS, nearly the entire optical interaction properties of a given chromophore may be captured in a single excitation event. The system was explored in freshly resected tissues, formalin fixed tissues, FFPE tissue sections, and histochemical stains. The multi-contrast technique was able to provide a convincing analogue to structures highlighted by traditional H&E-stained preparations, thus providing effective label-free H&E visualizations. Moreover, the TA-PARS captures a relative absorption contrast factor proposed as the QER. The QER yields an absorption metric, providing contrast which is not afforded by radiative or non-radiative absorption techniques independently. This report represents an exciting exploration into the potential new landscape of label-free microscopic inspection of biological media.

**Acknowledgements:**

The authors, thank the staff of the University of Waterloo Central Animal Facility, specifically Jean Flanagan for her work in procuring animal tissue specimens. Without her generous contribution this would not be possible.

The authors thank the following sources for funding used during this project. Natural Sciences and Engineering Research Council of Canada (DGECR-2019-00143, RGPIN2019-06134); Canada Foundation for Innovation (JELF #38000); Mitacs Accelerate (IT13594); University of Waterloo Startup funds; Centre for Bioengineering and Biotechnology (CBB Seed fund); illumiSonics Inc (SRA #083181); New frontiers in research fund – exploration (NFRFE-2019-01012).

**Author Contributions:**

Benjamin Ecclestone collected results, prepared the figures, and wrote the main manuscript. Dr. Kevan Bell assisted in collecting results and writing the manuscript. Sarah Sparkes assisted in preparing results and figures. Dr. Deepak Dinakaran and Dr. John Mackey worked to collect and prepare tissue specimens and acted as clinical consultation in the assessment of results. Dr. Parsin Haji Reza directed and organized the project, and oversaw experimental work, and manuscript writing as the principal investigator.

**Disclosures:**

Authors Benjamin Ecclestone, Kevan Bell, Deepak Dinakaran, John R. Mackey and Parsin Haji Reza, all have financial interest in IllumiSonics which has provided funding to the PhotoMedicine Labs.

# Supplemental Information

*Section 1: Contrast Mechanisms of Popular Microscopy Techniques*

**Table. 1** Popular microscopy techniques and their relevant contrast mechanisms.

| Modality | Radiative Absorption | Non-Radiative Absorption | Scattering |
|---|---|---|---|
| Photoacoustic Remote Sensing (PARS) | ✗ | ✓ | ✓ |
| Total Absorption-PARS (TA-PARS) | ✓ | ✓ | ✓ |
| Photoacoustic microscopy | ✗ | ✓ | ✗ |
| Confocal Microscopy | ✗ | ✗ | ✓ |
| Fluorescence | ✓ | ✗ | ✓ |
| Multiphoton Fluorescence | ✓ | ✗ | ✓ |
| Stimulated Raman Scattering Microscopy | ✓ | ✗ | ✓ |
| Optical Coherence Tomography | ✗ | ✗ | ✓ |

*Section 2: Comparison of the TA-PARS visualizations with previous PARS embodiments*

**Table. 2** System characteristics of previously reported PARS systems featuring comparable 266 nm excitation sources. Note, "-" indicates that the relevant value was not reported in the given study.

| Excitation Wavelength (nm) | Excitation Energy (nJ) | Detection Wavelength (nm) | Detection Power (mW) | Lateral Res. (µm) | Axial Res. (µm) | Citation |
|---|---|---|---|---|---|---|
| 266 | - | 1310 | - | 1.2 | 7.3 | 1 |
| 266 | 10 | 1310 | 10 | 0.69 | - | 2 |
| 266 | 3 | 1310 | - | 1.2 | - | 3 |
| 532 | 15 | 1310 | - | 1.5 | - | 3 |
| 250 | 0.9-20 | 1310 | - | 0.58 | - | 4 |
| 266 | 0.9-20 | 1310 | - | 2.5 | - | 4 |
| 405 | 0.9-20 | 1310 | - | 0.59 | - | 4 |
| 266 | - | 1310 | - | 0.425 | - | 5 |
| 266 | - | 1310 | - | - | - | 6 |
| 266 | 0.75 | 1310 | - | 0.3 | - | 7 |
| 266 | 10 | 1310 | 7.8 | 0.39 | 1.2 | 8 |
| 266 | 5 | 1310 | - | 0.39 | 1.2 | 9 |
| 266 | 5 | 1310 | 7 | 0.44 | - | 10 |
| 532 | 80 | 1310 | 7 | 1.2 | - | 10 |
| 1225 | 485 | 1550 | 15 | 0.96 | 17.4 | 11 |
| 266 | 3 | 1310 | 3.24 | 0.39 | - | 12 |
| 266 | 10 | 1310 | 7.76 | 0.39 | - | 12 |

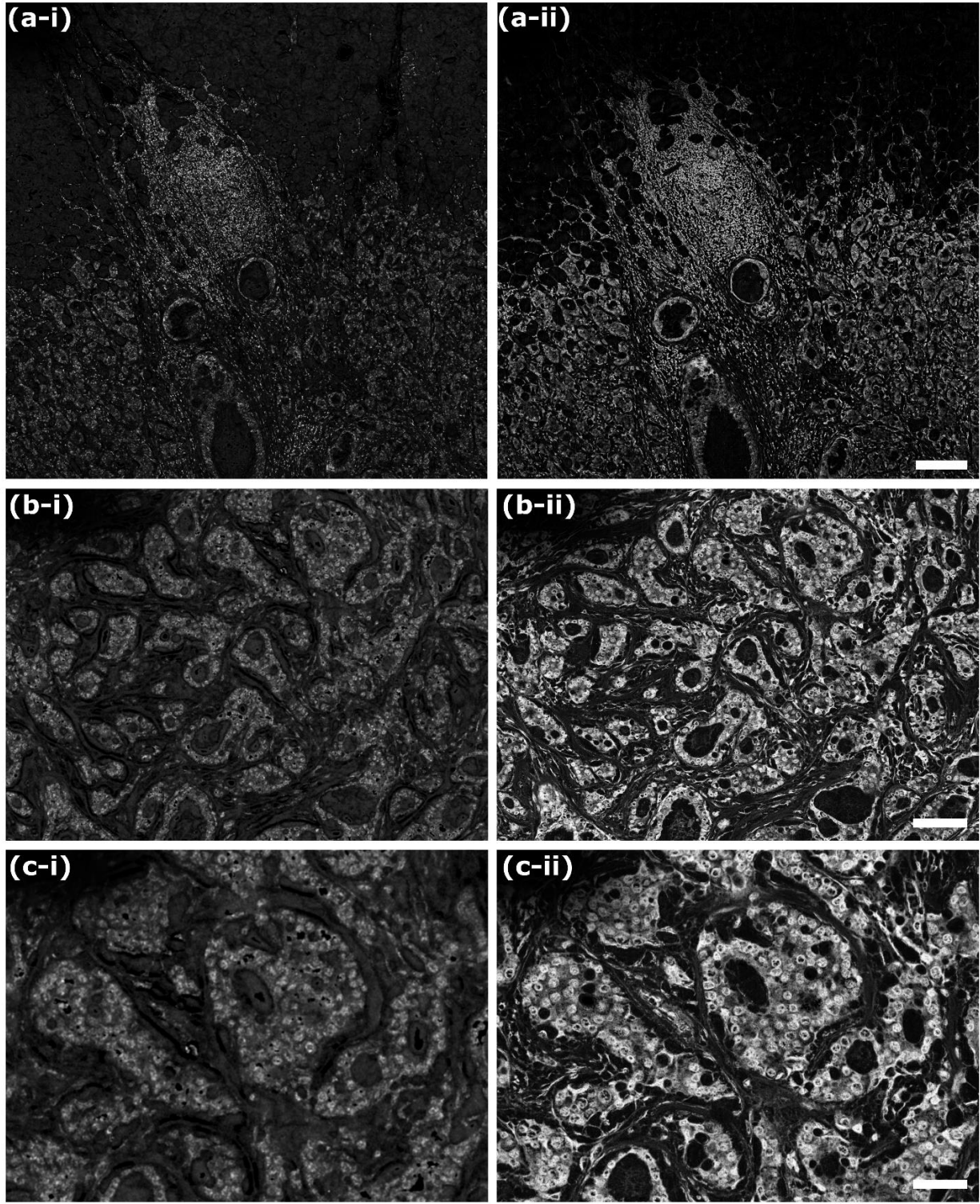

**Fig. 1** Comparison of the TA-PARS visualizations in resected tissues to the previously reported PARS system, used by Ecclestone et al.[7]. (a) Wide field of view comparison (a-i) First generation PARS (a-ii) TA-PARS. Scale Bar: 200 µm. (b) Midfield comparison (b-i) First generation PARS (b-ii) TA-PARS. Scale Bar: 100 µm. (c) Small field of view close up comparison (c-i) First generation PARS (c-ii) TA-PARS. Scale Bar: 200 µm.

*Section 3: Characterization of the TA-PARS imaging performance*

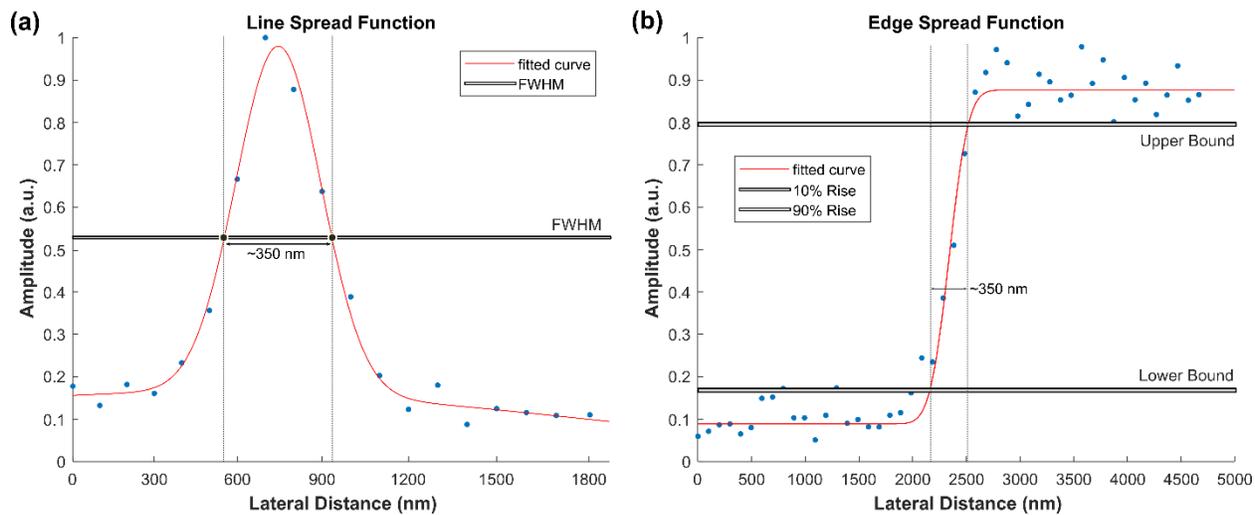

**Fig. 2** TA-PARS resolution measurement examples. TA-PARS resolution was determined to be ~350 nm based on the average (a) line spread function generated from imaging sub resolution fibers, and (b) the edge spread function generated imaging nuclei.

*Section 4: TA-PARS images*

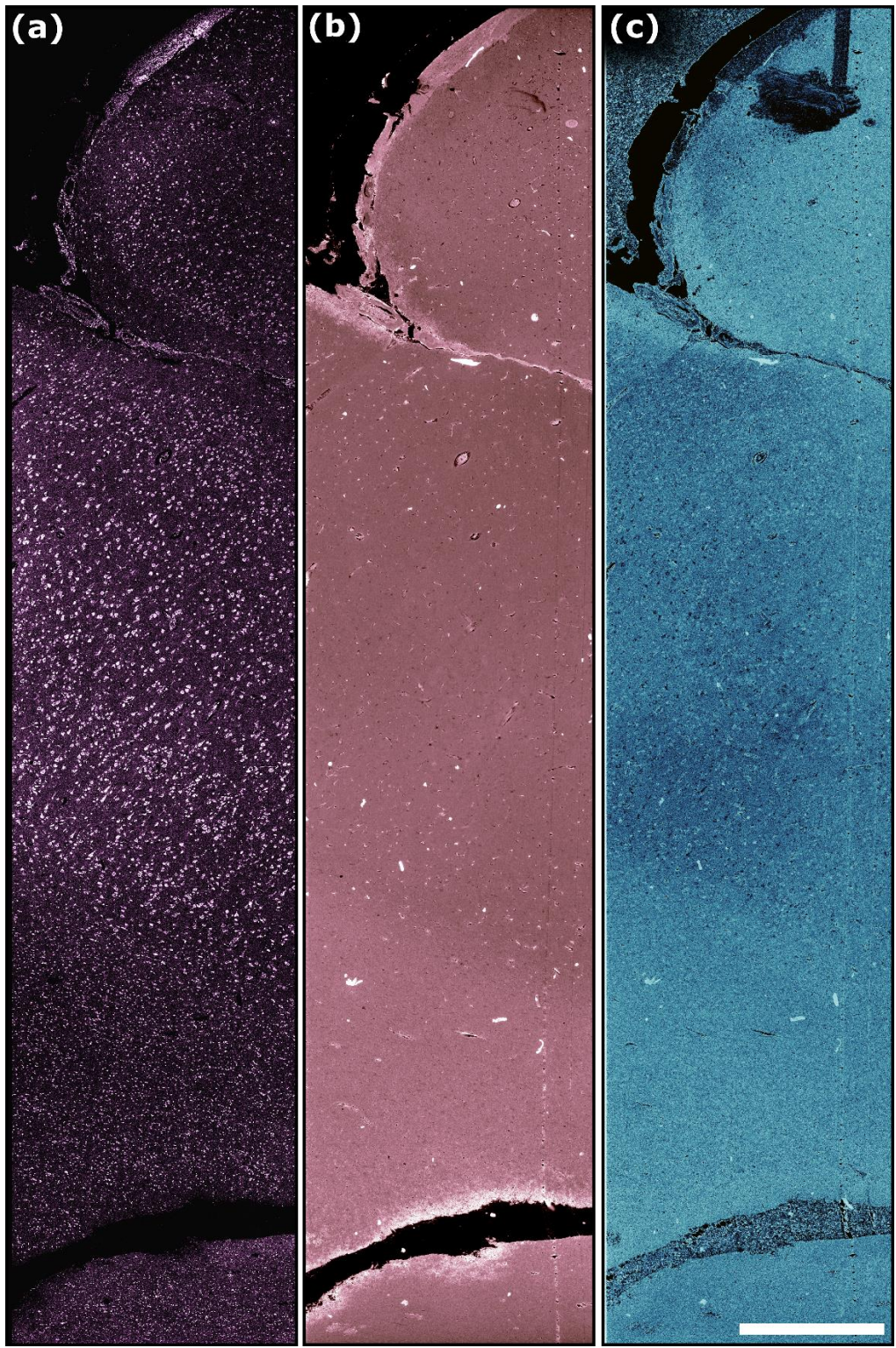

**Fig. 3** Different contrasts afforded by the TA-PARS microscope in a thin section of formalin fixed paraffin embedded human brain tissues. (a) non-radiative absorption contrast (b) radiative absorption contrast (c) optical scattering. contrast Scale Bar: 1 mm

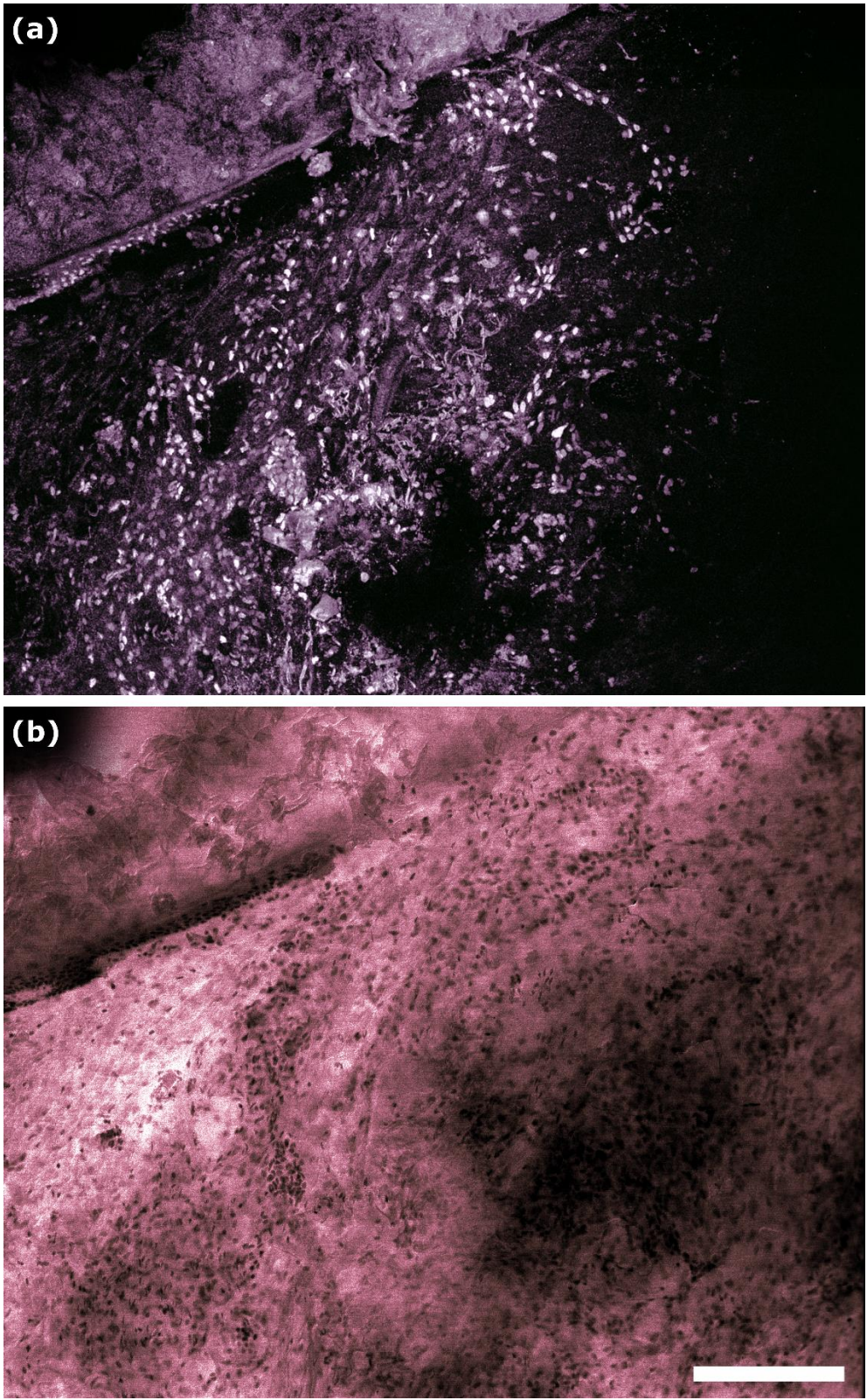

**Fig. 4** TA-PARS images in bulk unprocessed resected human skin tissues. (a) Non-radiative absorption contrast. (b) Radiative absorption contrast. Scale Bar: 200 μm

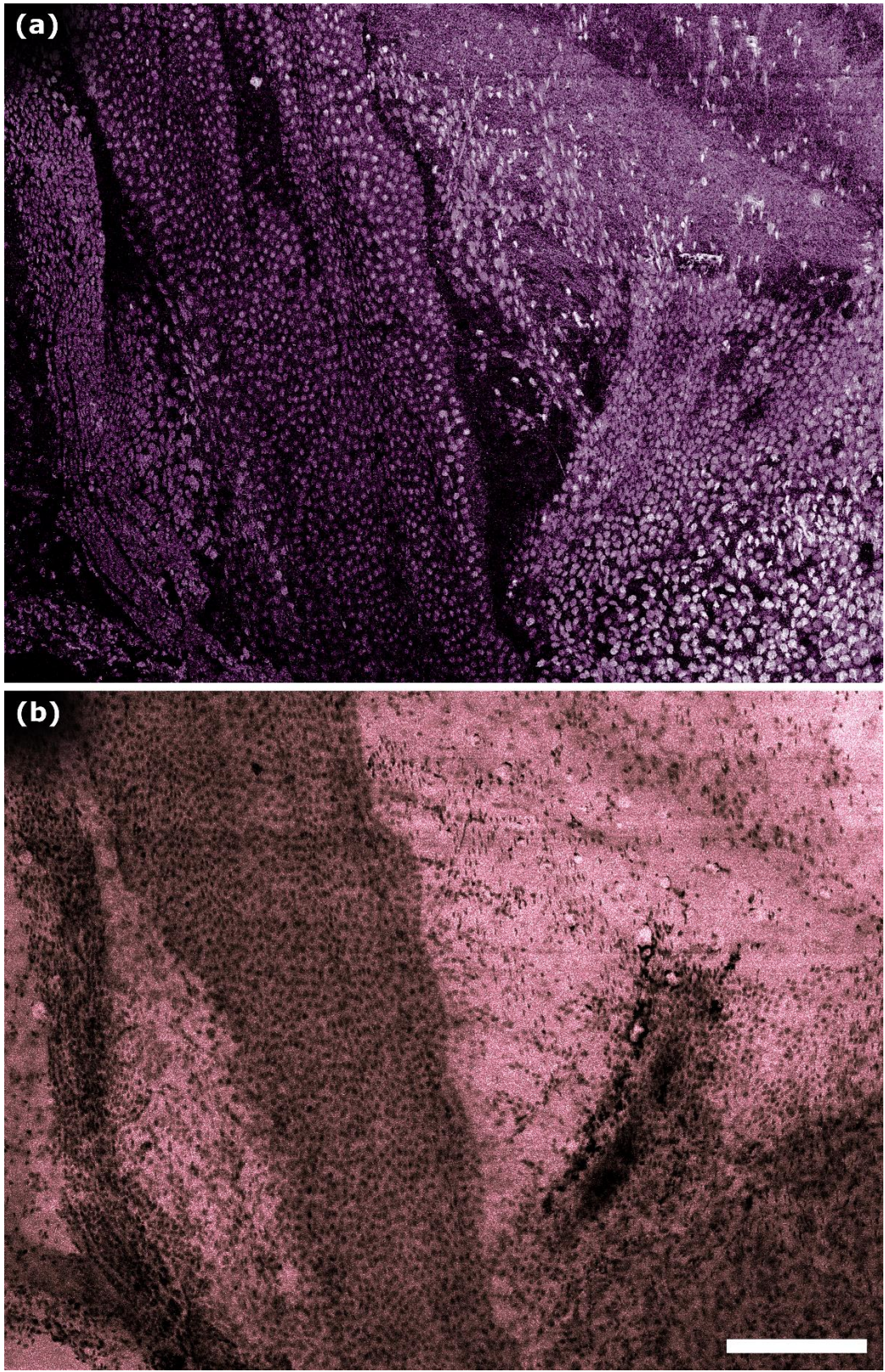

**Fig. 5** TA-PARS images in sections of resected Rattus brain tissues. (a) Non-radiative absorption contrast. (b) Radiative absorption contrast. Scale Bar: 200 μm

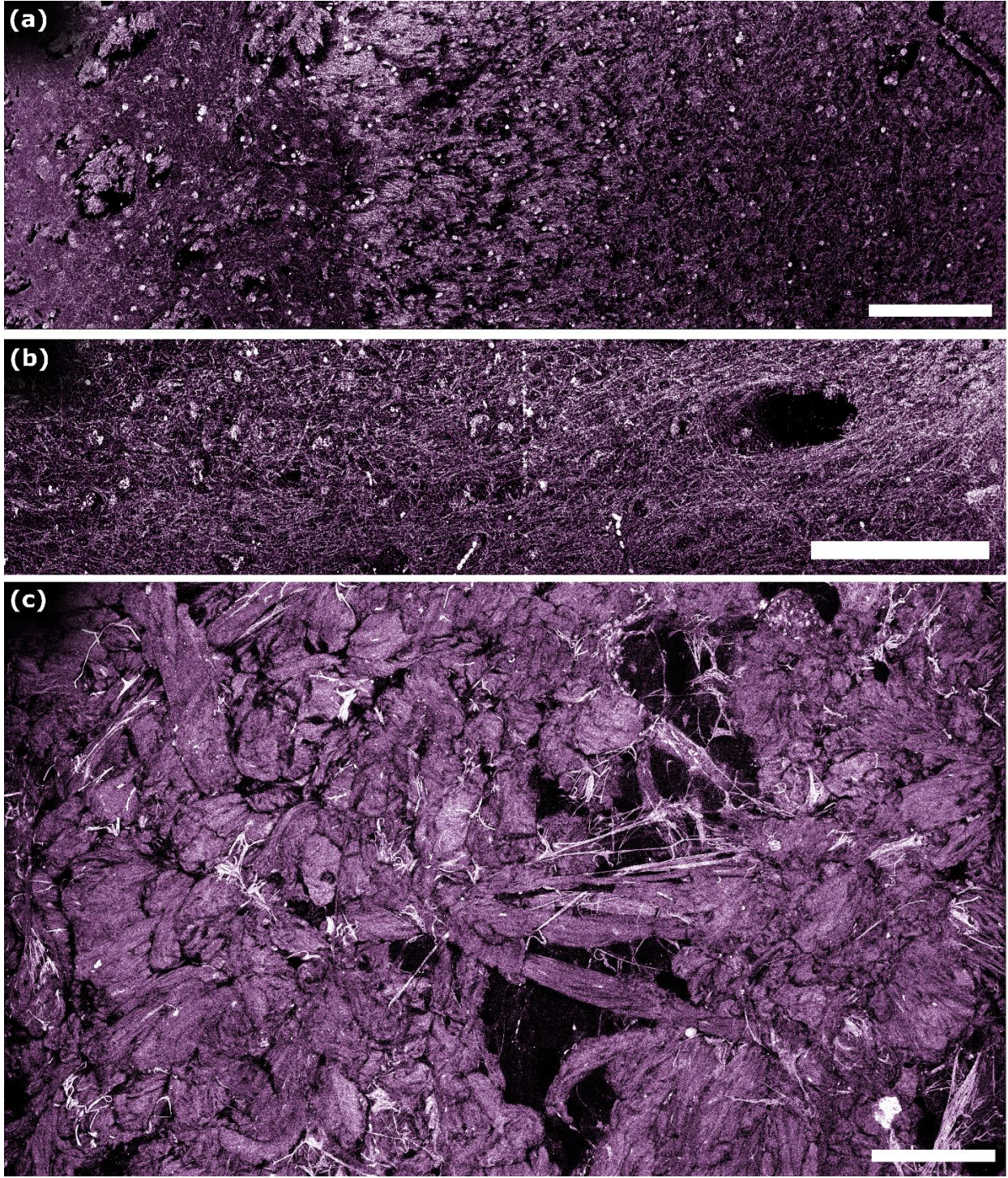

**Fig. 6** TA-PARS non-radiative absorption contrast images in bulk resected tissue specimens. (a) Rattus brain tissues showing the boundary between two tissue regions. Scale Bar: 200 µm. (b) Murine brain tissues exhibiting interwoven neuron structures and sparse nuclei. Scale Bar: 200 µm. (c) Subcutaneous human skin tissues exhibiting collagen, lipids, and fibrin structures. Scale Bar: 200 µm.

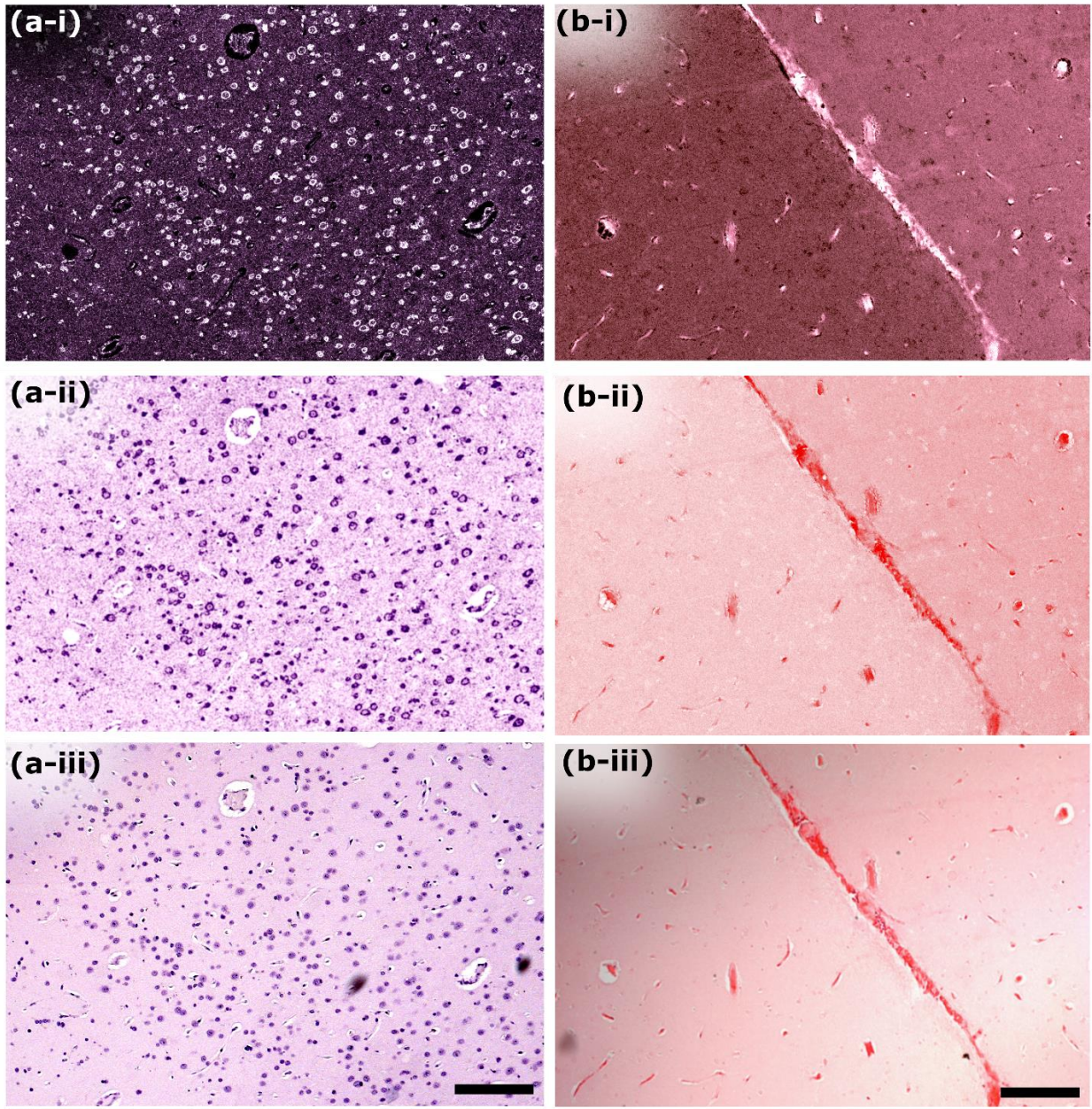

**Fig. 7** Comparison of the TA-PARS non-radiative and radiative contrast to hematoxylin and eosin staining, respectively. (a) Comparison of TA-PARS non-radiative contrast with hematoxylin staining (a-i) TA-PARS non-radiative absorption contrast (a-ii) TA-PARS non-radiative absorption contrast, with emulated hematoxylin color mapping (a-iii) The same section of tissues, stained with hematoxylin stain, and imaged with a brightfield microscope. (b-i) TA-PARS radiative absorption contrast (b-ii) TA-PARS radiative absorption contrast, with emulated hematoxylin color mapping (b-iii) The same section of tissues, stained with eosin stain, and imaged with a brightfield microscope. d

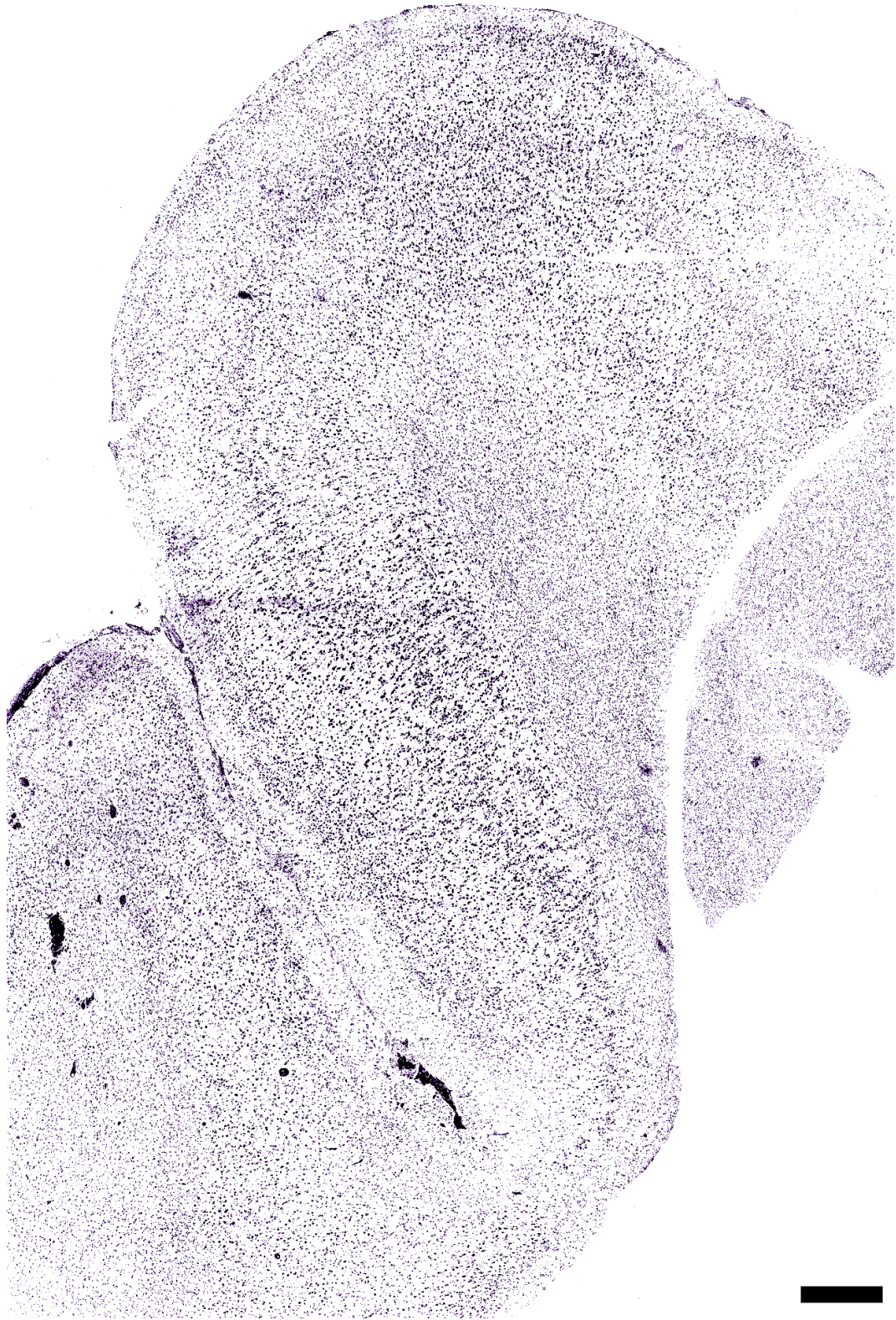

**Fig. 8** TA-PARS non-radiative contrast imaging of nearly an entire thin section of FFPE human brain tissues, providing visualizations analogous to hematoxylin staining. Image is artificially colorized to represent hematoxylin staining contrast. Scale Bar: 2 mm.

.

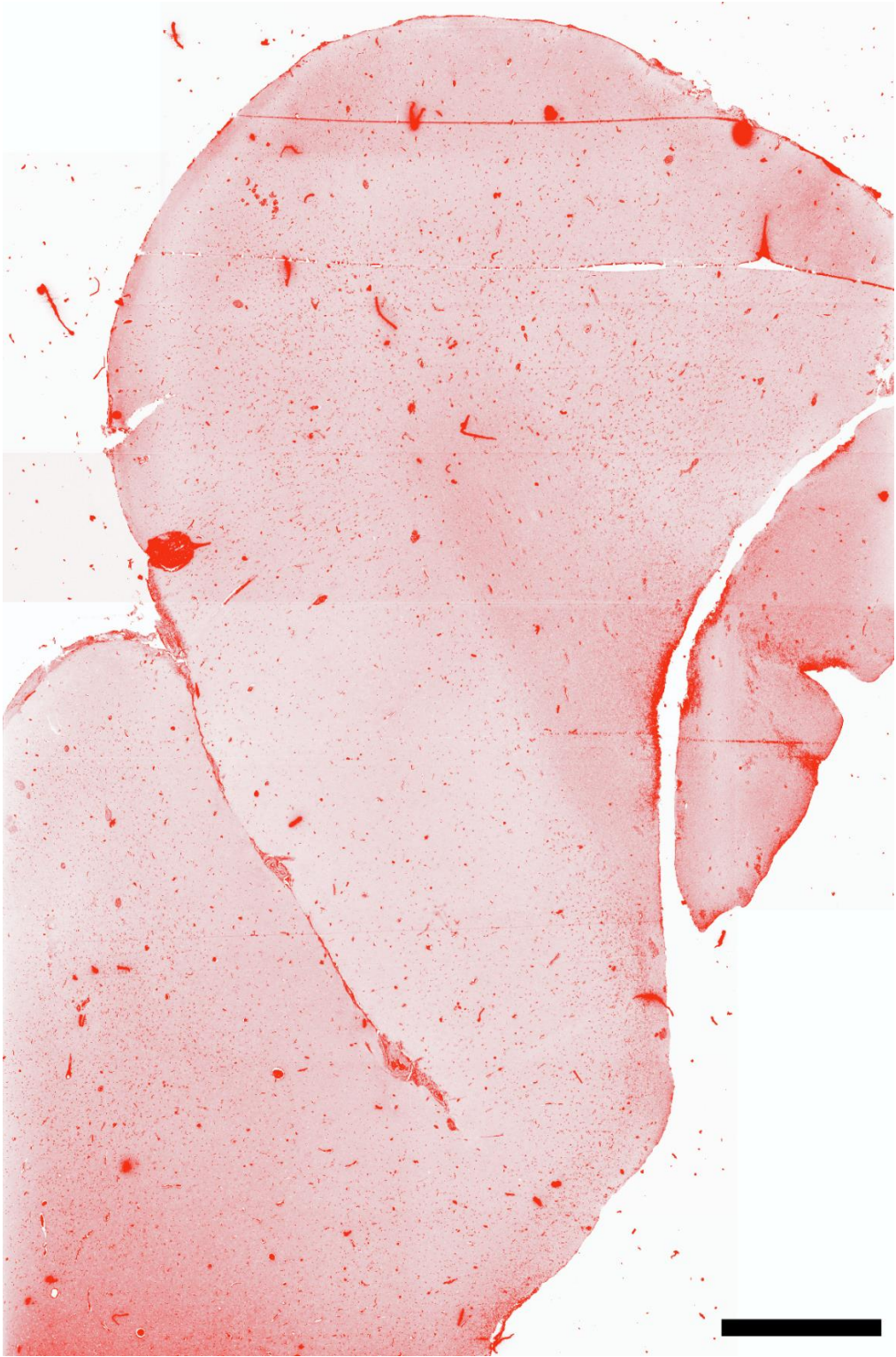

**Fig. 9** TA-PARS radiative contrast imaging of nearly an entire thin section of FFPE human brain tissues providing visualizations analogous to eosin staining. Image is artificially colorized to represent eosin staining contrast. Scale Bar: 2 mm.

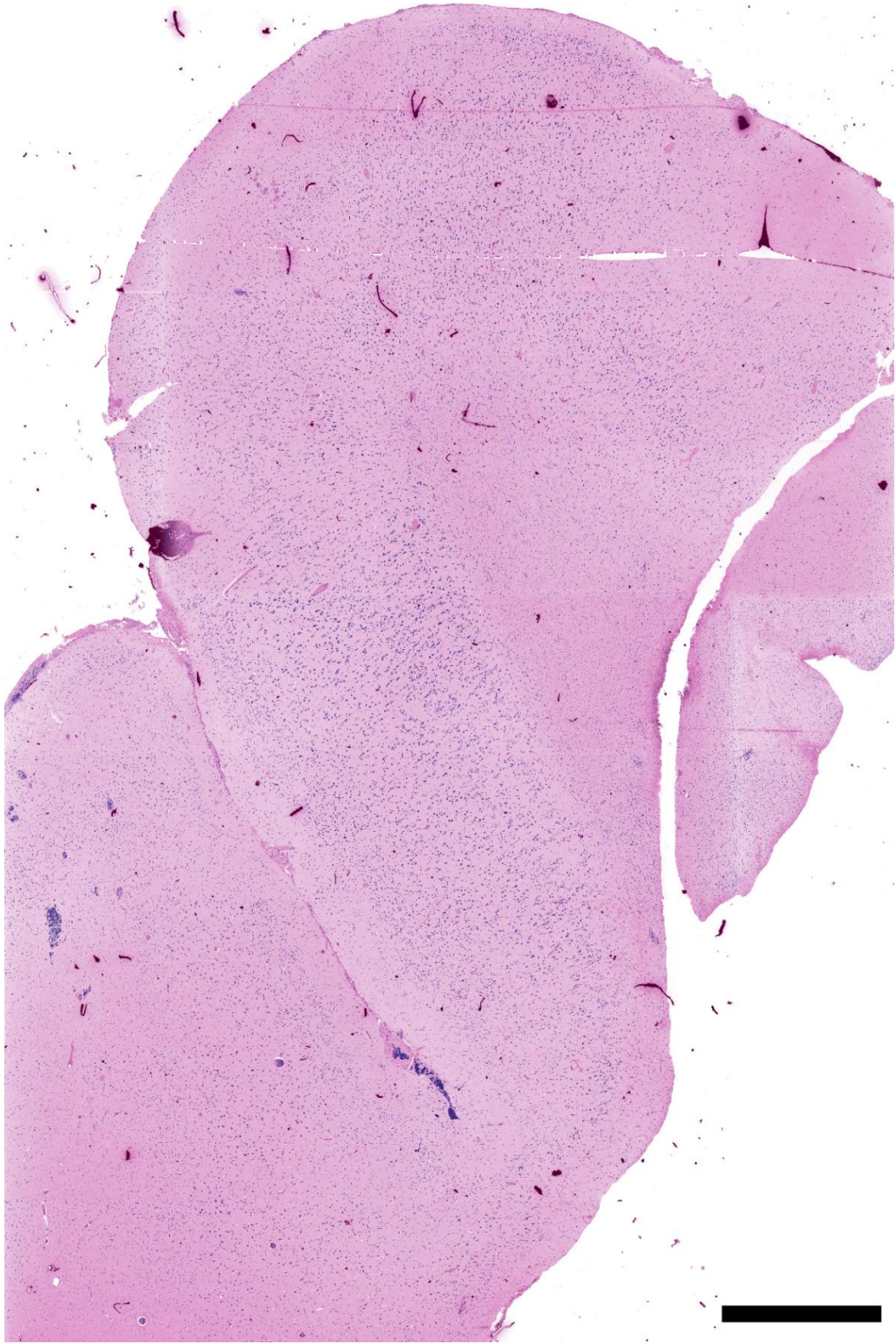

**Fig. 10** TA-PARS emulated H&E staining of nearly an entire thin section of FFPE human brain tissues. Scale Bar: 2 mm.

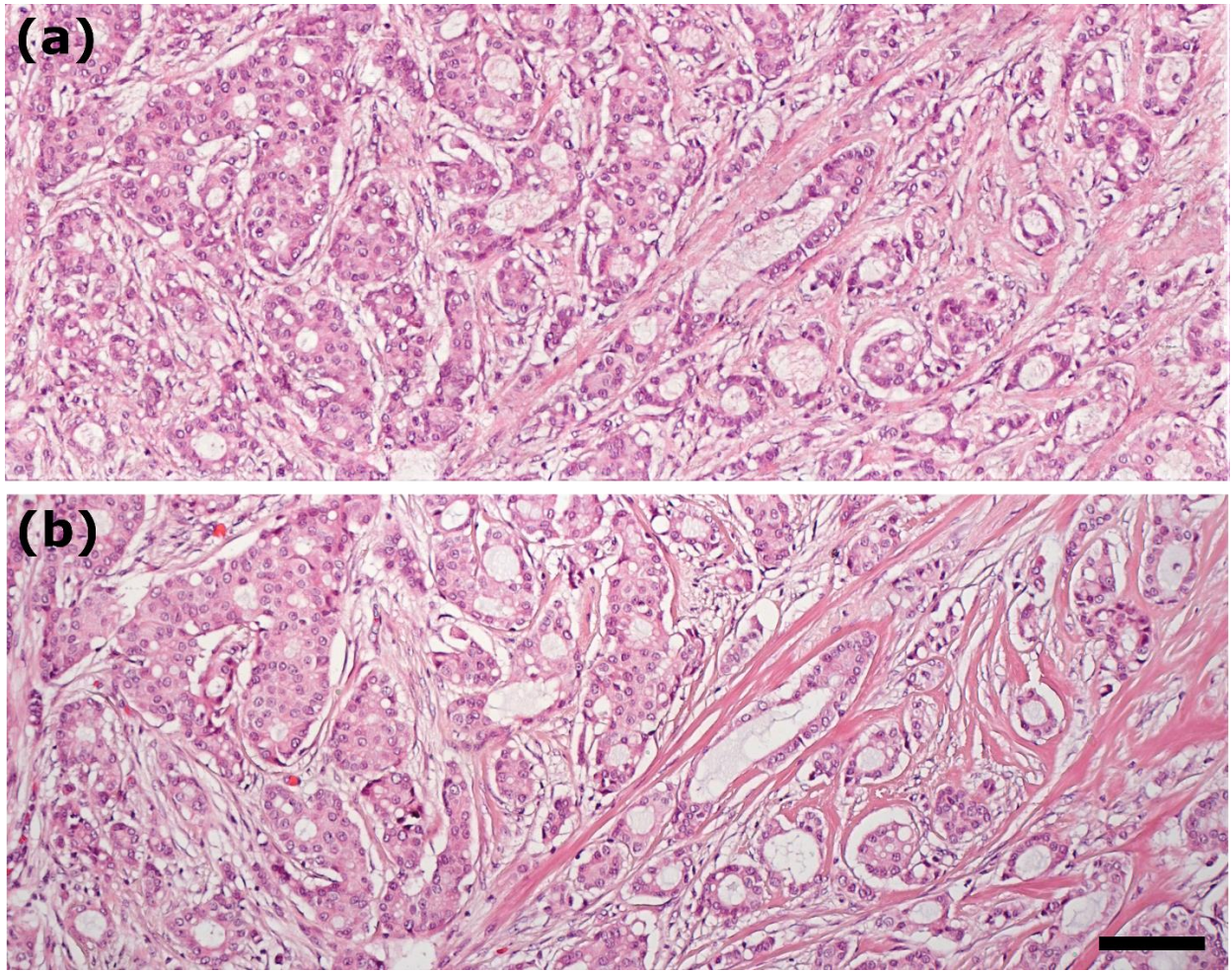

**Fig. 11** One-to-one comparison of TA-PARS emulated H&E staining and traditional H&E staining in thin sections of resected human breast tissues. (a) TA-PARS Emulated H&E image (b) Same section of tissues imaged under a brightfield microscope following H&E staining. Scale Bar: 100 µm.

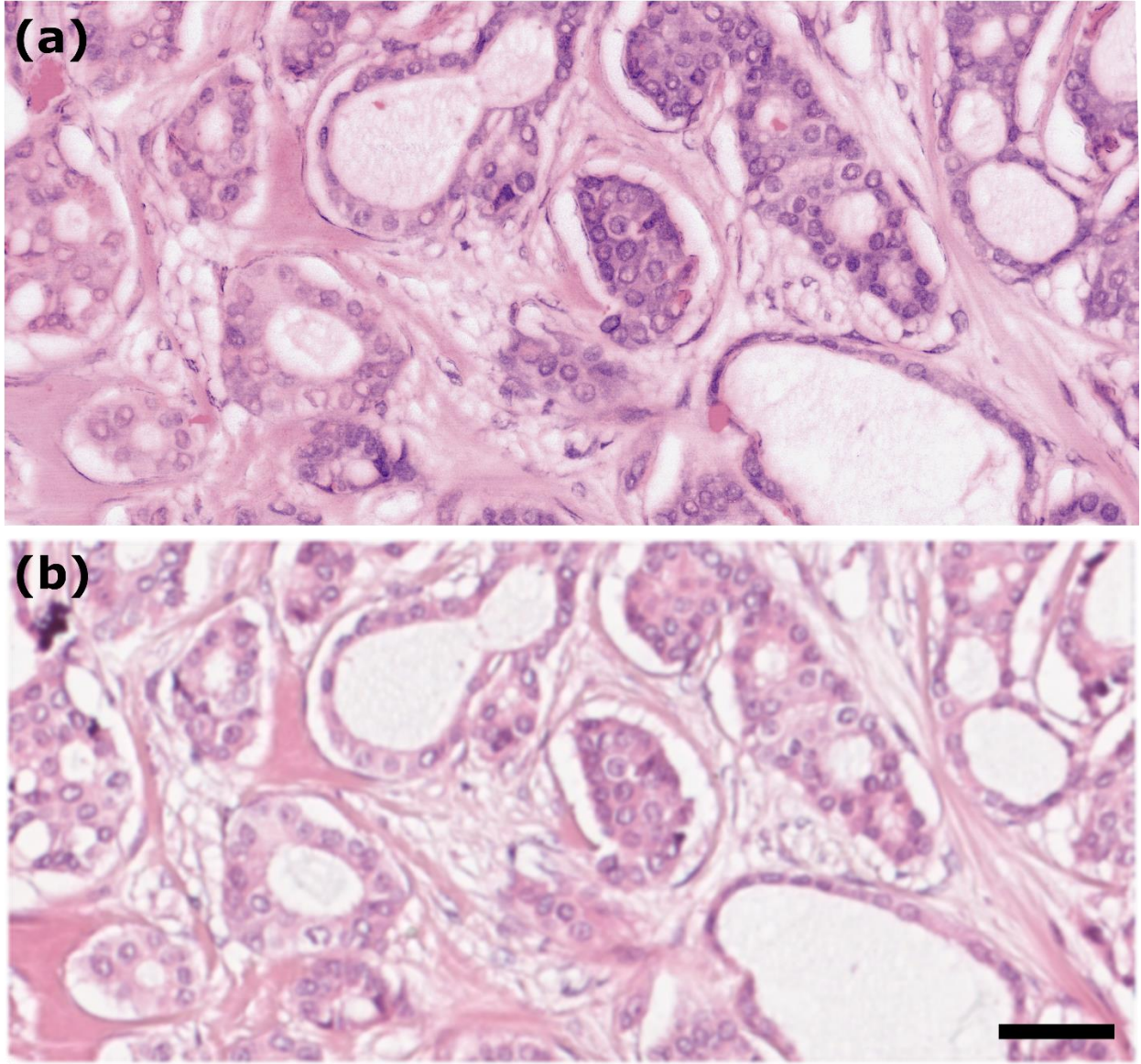

**Fig. 12** One-to-one comparison of TA-PARS emulated H&E staining and traditional H&E staining in thin sections of resected human breast tissues. (a) TA-PARS Emulated H&E image (b) Same section of tissues imaged under a brightfield microscope following H&E staining. Scale Bar: 50 µm.

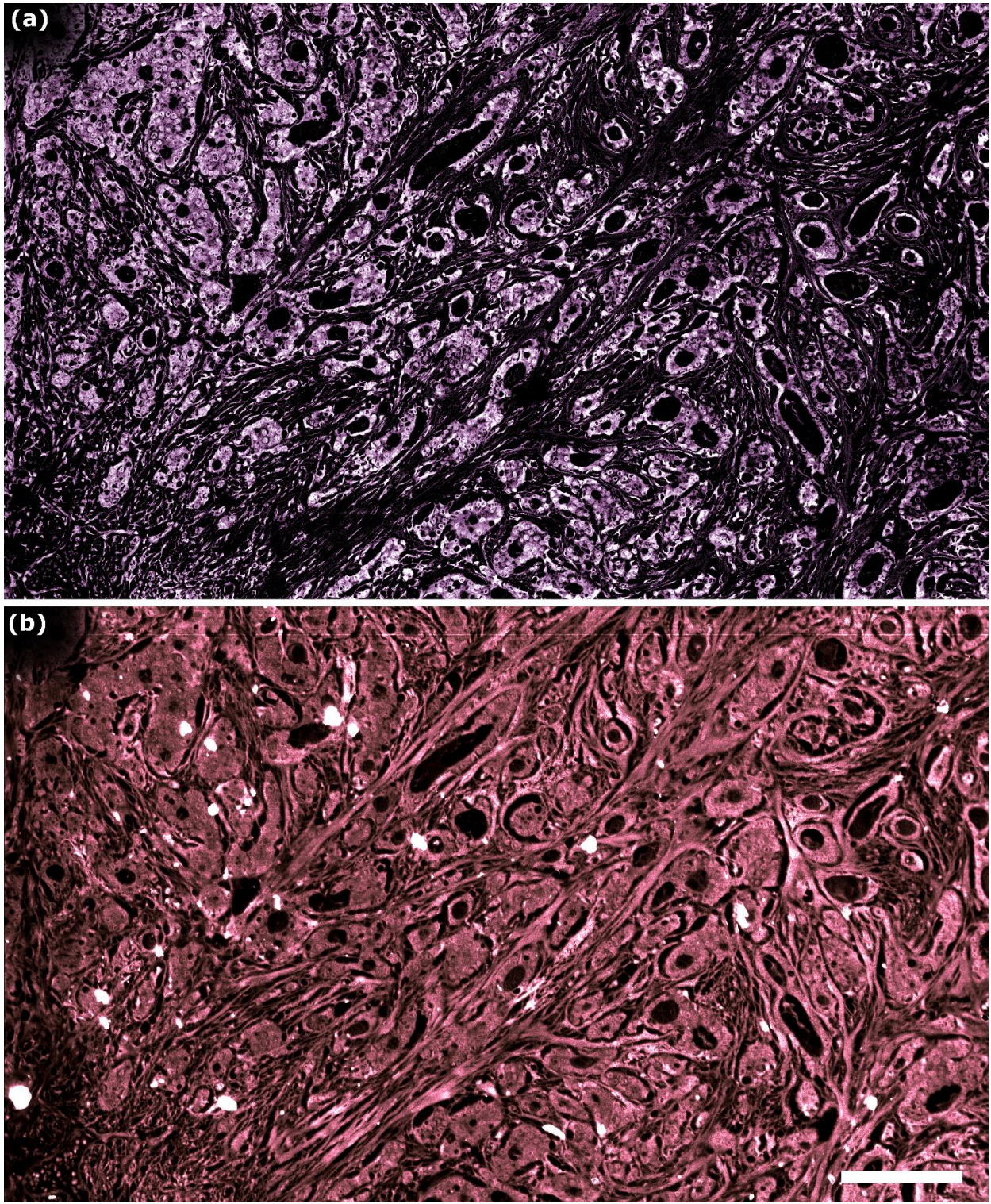

**Fig. 12** TA-PARS images in thin sections of FFPE human breast tissues. (a) Non-radiative absorption contrast. (b) Radiative absorption contrast. Scale Bar: 200 µm.

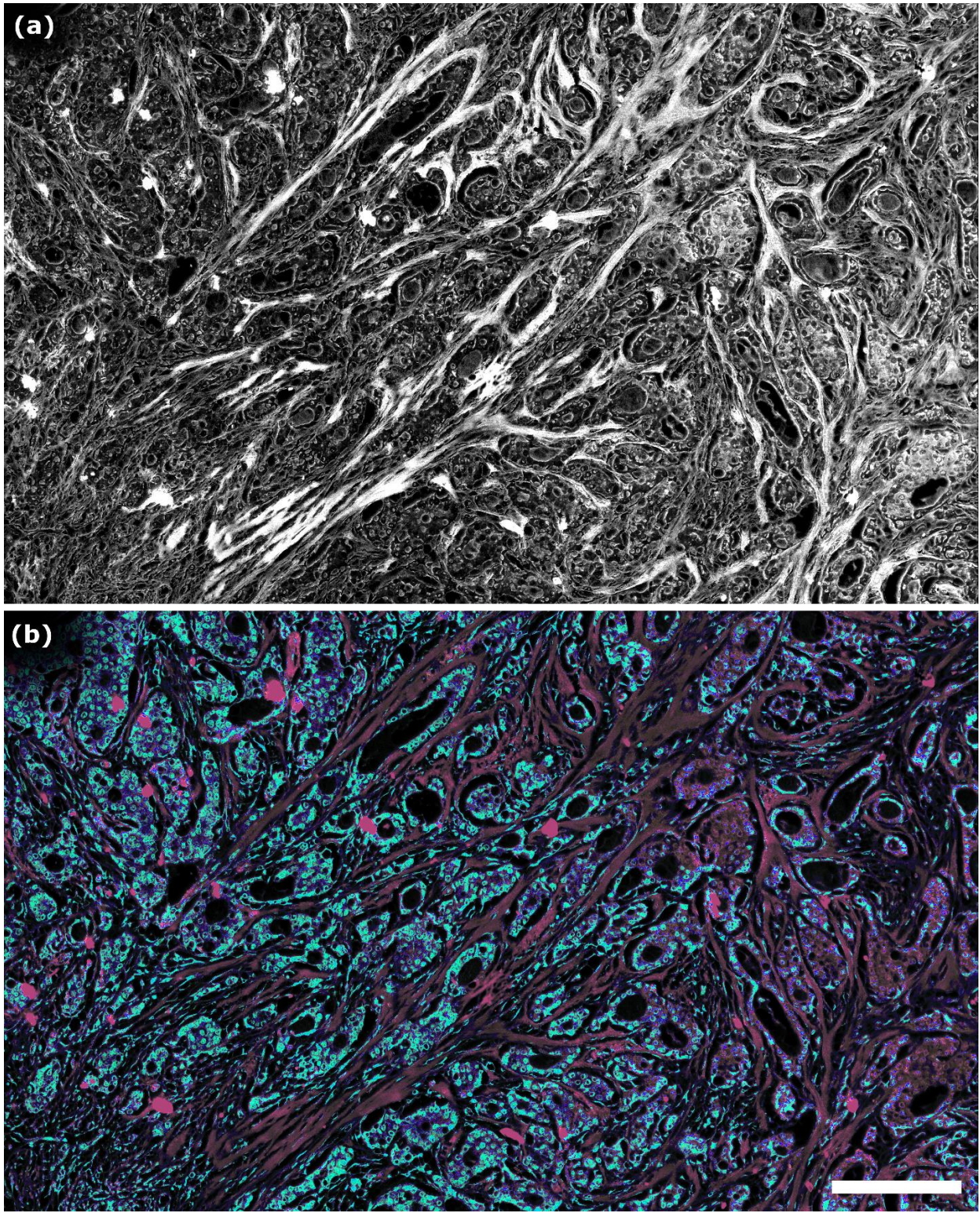

**Fig. 13** TA-PARS quantum efficiency ratio imaging (QER) in thin sections of FFPE human breast tissues. (a) Gray scale QER visualization. (b) False color representation of the QER, where the color is defined by the QER, and intensity is defined by the total-absorption level. Scale Bar: 200 µm.